\documentclass[aps,pra,superscriptaddress,reprint]{revtex4-1}

\usepackage{amsmath}
\usepackage{amsfonts}
\usepackage{graphicx}
\usepackage{epstopdf}
\usepackage{bm}
\usepackage{color}

\renewcommand{\textcolor}[2]{#2}

\begin{document}

\bibliographystyle{apsrev}

\title{Coherence in parametric fluorescence}

\author{T. Onodera}
\affiliation{Department of Physics and Institute for Optical Sciences, University of Toronto, 60 St.
George St., Toronto, ON M5S1A7, Canada}

\author{Marco Liscidini}
\affiliation{Dipartimento di Fisica, Universit\`{a} degli Studi di Pavia, via Bassi 6, Pavia, Italy}

\author{J. E. Sipe}
\affiliation{Department of Physics and Institute for Optical Sciences, University of Toronto, 60 St.
George St., Toronto, ON M5S1A7, Canada}

\author{L. G. Helt}
\affiliation{Centre for Ultrahigh bandwidth Devices for Optical Systems (CUDOS),
QSciTech Research Centre, MQ Photonics Research Centre,
Department of Physics and Astronomy, Macquarie University, NSW 2109, Australia}

\date{\today}

\begin{abstract}
We investigate spontaneous four wave mixing (SFWM) in a single-channel side-coupled integrated spaced sequence of resonators (SCISSOR). Analytic expressions for the number of photon pairs generated, as well as the biphoton wave function (joint spectral amplitude) describing the pairs, are derived and numerically computed for different pump pulse durations and numbers of ring resonators. In the limit of a long input pump pulse, we show a strong analogy between super-linear scaling of generation efficiency with respect to the number of rings in the structure and Dicke superradiance. More generally, we discuss in detail the factors that influence the shape of the biphoton wave function, as well as the conditions for observing super-SFWM. 
\end{abstract}


\maketitle

\section{Introduction}

The last few years have witnessed a burgeoning interest in integrated quantum nonlinear optics \cite{lanco06,takesue07,clemmen09,xiong11,davanco12,azzini12,mookherjea13,ducci13}. Nonlinear optical processes, such as spontaneous parametric down conversion (SPDC) and spontaneous four-wave-mixing (SFWM), have been explored as reliable and efficient sources of nonclassical states of light in integrated semiconductor structures. The development of such sources is one of the major challenges in implementing a full integration of quantum optics on-a-chip \cite{galland13}, which holds the promise of a host of applications ranging from quantum information processing to quantum computing. There is also an interest in integrated quantum optics beyond developing technological applications, as quantum optics has always been an important resource for the study of fundamental concepts in quantum mechanics, including entanglement \cite{kwiat95} and weak measurement \cite{steinberg09}. More recently, integrated optics has been used to demonstrate fundamental results such as Anderson localization in the quantum regime \cite{mataloni13} and boson sampling \cite{walmsley13,white13,tilmann13}.

Both SPDC and SFWM have corresponding classical processes, respectively difference frequency generation and stimulated four-wave-mixing. Most of the conditions that lead to high efficiency for these classical phenomena are those required to achieve high efficiency for the corresponding quantum processes \cite{helt12,azzini12_2,azzini13}. Thus, the design of integrated devices to enhance the nonlinear light-matter interaction in the quantum regime can take inspiration from recent results concerning classical frequency conversion in integrated devices \cite{absil00,vlasov06,yang07,turner08,corcoran09,galli10}, and from the know-how developed over more than fifty years of classical nonlinear optics. Similarly, it makes sense to investigate parametric fluorescence in devices that have been initially proposed for applications in linear optics, as many of them have peculiar properties that could be exploited in the quantum nonlinear regime. In part, this was the motivation for our recent work on SFWM in single-channel side-coupled integrated spaced sequences of resonators (SCISSORs) \cite{helt12_2}. Known for years as linear components in integrated optics \cite{heebner02,heebner04}, we showed that these systems have interesting properties in the nonlinear quantum optics regime: for continuous wave (CW) pump excitation, the probability of photon pair production by SFWM can scale quadratically with the number of resonators, a phenomenon we named super-SFWM.

\textcolor{red}{Superradiance, or the phenomenon of several emitters acting in a coherent way that leads to an enhanced emission rate, was first explored by Dicke~\cite{dicke54} in his classic paper ``Coherence in Spontaneous Radiation Processes,'' in relation to the spontaneous emission of a gas of atoms.  In the present work we cast superradiance in a more general light, and consider super-SFWM as an instance of quantum nonlinear optical processes acting coherently.  While such coherence is quite general, and can be observed for SFWM in a variety of structures, here we use the simple example of a SCISSOR, which we believe best illustrates the connection between the superradiance of super-SFWM and that of Dicke's work.  Extending our earlier work on generation efficiency with CW pumping~\cite{helt12_2}, here we formalize the aforementioned connection, relax the CW approximation, and explore the spectral properties of the state of generated photons in addition to its generation efficiency.}

\textcolor{red}{We begin in Section~\ref{sec:superradiance} with a simple calculation based on Fermi's Golden Rule. While this is strictly limited to identifying the rate at which pairs are generated, it has the advantage of explicitly showing a parallel between coherence in parametric fluorescence and coherence in spontaneous emission.  We start with a general discussion of the simple theory of superradiance, and identify the parallels between Dicke's example of atoms in a gas and the example of rings in a SCISSOR structure.  This simple calculation shows that super-SFWM can arise from the constructive interference of the probability amplitude of generating a photon pair in each ring.  In Section~\ref{sec:longpulse} we show how, in the limit of a long pump pulse, general expressions for SFWM in a SCISSOR reduce to the simpler Fermi's Golden Rule result of Section~\ref{sec:superradiance}; details of the application of the asymptotic fields formalism~\cite{liscidini12} to a SCISSOR, the long input pump pulse limit applied to these expressions, manipulations involved in the connection with Dicke superradiance, and the derivation of the maximum number of rings that are expected to behave coherently are relegated to Appendices  \ref{sec:formalism}, \ref{sec:longpulsedetails}, \ref{sec:connection}, and \ref{sec:coherencenumber}, respectively.}

Armed with a general description of SFWM in a SCISSOR, even in the presence of pump pulses with duration comparable or shorter than the dwelling time of a photon in each ring resonator, we go beyond the long pulse limit in Section \ref{sec:coherence} and study the time-energy correlation of the generated photon-pairs in general. We evaluate the biphoton wave function (joint spectral amplitude) that characterizes them, revealing how the interference depends both on the spectral features of the pump pulse, and on the geometrical parameters of the structure. In particular, this suggests a strategy for the design of \textit{ad hoc} biphoton wave functions, in which the quantum correlations of the generated photon pairs are tailored by adjusting the interference between different sources driven by the same pump pulse. We examine how the generation rate of photon pairs depends on the properties of the pump pulse, and particularly how the number of rings that radiate coherently depends on that pump pulse. We also present some numerical results for a specific structure, and illustrate the biphoton wave functions that result as the number of rings and the pump pulse duration are both varied. This allows us to identify the conditions that must be satisfied to observe super-SFWM in a given structure, and how they depend on geometrical and material parameters. \textcolor{red}{Some considerations, and connections to other third-order nonlinear optical processes, are discussed in Section~\ref{sec:discussion}, and} our conclusions are presented in Section \ref{sec:conclusions}.

\section{Superradiance: atoms and rings\label{sec:superradiance}}

To ground our analysis of super-SFWM in something familiar, we begin with a simple Fermi's Golden Rule calculation that establishes a connection between super-SFWM and Dicke superradiance. Discussions of superradiance typically involve an output, or ``radiation'' subsystem with Hilbert space $\mathcal{H}_{\text{R}}$, consisting of at least part of the radiation field, and a ``pump'' subsystem, consisting of $N$ identical or nearly identical components, with Hilbert space $\mathcal{H}_{\text{P}}$. The full Hilbert space is a direct product of these two subsystem Hilbert spaces,
\begin{equation}
\mathcal{H}=\mathcal{H}_{\text{R}}\otimes\mathcal{H}_{\text{P}}.
\end{equation}
In Dicke's work \cite{dicke54} the output subsystem is the full radiation field and its initial state is the vacuum, while the pump subsystem is a collection of atoms and the initial state is an energy eigenstate. We slightly generalize the usual elementary treatment of superradiance by considering an initial state of the form
\begin{equation}
\left\vert \psi(0)\right\rangle =\left\vert \text{vac}\right\rangle \otimes\left\vert \Psi_{\text{P}}\right\rangle ,\label{initial}
\end{equation}
where $\left\vert \text{vac}\right\rangle $ indicates the vacuum state of the radiation subsystem, and $\left\vert \Psi_{\text{P}}\right\rangle$ is now an arbitrary initial state of the pump subsystem; for the moment we do not consider how the components of the pump subsystem were initialized in the state $\left\vert \Psi_{\text{P}}\right\rangle $. The Hamiltonian for the system can be written very generally as
\begin{equation}
H=H_{\text{R}}+H_{\text{P}}+V,\label{Htotal}
\end{equation}
where $H_{\text{R}}$ and $H_{\text{P}}$ are respectively the Hamiltonians of the isolated radiation and pump systems, and $V$ describes the interaction between them. We will see that in our examples that the salient part of the interaction $V$ can be written as
\begin{equation}
V=-\sum_{m,q}c_{mq}\mathcal{R}_{q}^{\dagger}\mathcal{P}_{m}+\text{H.c.},\label{Vform}
\end{equation}
where the $c_{mq}$ are constants, $\mathcal{R}_{q}^{\dagger}$ is a raising operator acting on kets in the radiation Hilbert space, $\mathcal{P}_{m}$ is a destruction operator acting on kets in the pump Hilbert space, and H.c. denotes Hermitian conjugate. Here $m$ labels the component of the pump system, $m=1,2,\ldots,N$, and $q$ is a parameter characterizing the output system. For an initial state \eqref{initial} the ket will evolve to $\left\vert \psi(t)\right\rangle$ at time $t$, and we seek the probability $\mathfrak{P}_{r}(t)$ that an appropriate measurement performed on the radiation system would find it in an energy eigenstate $\left\vert \psi_{r}\right\rangle $ of $H_{\text{R}}$, independent of the state of the pump system,
\begin{equation}
\mathfrak{P}_{r}(t)=\sum_{p}\left\vert \left(\left\langle \bar{\psi}_{p}\right\vert \otimes\left\langle \psi_{r}\right\vert \right)\left\vert \psi(t)\right\rangle \right\vert ^{2},
\end{equation}
where the $\left\vert \bar{\psi}_{p}\right\rangle $ are energy eigenstates of $H_{\text{P}}$. Applying perturbation theory to calculate $\left\vert \psi(t)\right\rangle $, a calculation under the usual assumptions of Fermi's Golden Rule leads to the result
\begin{align}
\frac{\mathfrak{P}_{r}(t)}{t}=&\frac{2\pi}{\hbar}\delta(\hbar\omega_{r}-\hbar\Omega)\nonumber \\
& \times\sum_{m,m^{\prime}}\left\langle \Psi_{\text{P}}|\mathcal{Q}^{\dagger}\left(m,r\right)\mathcal{Q}\left(m^{\prime},r\right)|\Psi_{\text{P}}\right\rangle ,\label{FGR}
\end{align}
where $H_{\text{R}}\left\vert \psi_{r}\right\rangle =\hbar\omega_{r}\left\vert \psi_{r}\right\rangle $, we take the energy of the vacuum state of $H_{\text{R}}$ to vanish, and we have assumed $e^{iH_{\text{P}}t/\hbar}\mathcal{P}_{m}e^{-iH_{\text{P}}t/\hbar}=\mathcal{P}_{m}e^{-i\Omega t}$ for some frequency $\Omega$; this can easily be generalized, but it will suffice for the examples of interest here. The operators $\mathcal{Q}\left(m,r\right)=\sum_{q}c_{mq}\left\langle \psi_{r}|\mathcal{R}_{q}^{\dagger}|\text{vac}\right\rangle \mathcal{P}_{m}$ are operators over the pump Hilbert space that depend parametrically on the final state $\left\vert \psi_{r}\right\rangle$ of the radiation system being considered; it is understood that \eqref{FGR} is to be summed over all possible final states $\left\vert \psi_{r}\right\rangle $, with that sum then passing to an integral in the continuum limit. Finally, we write \eqref{FGR} as
\begin{equation}
\frac{\mathfrak{P}_{r}(t)}{t}=\frac{2\pi}{\hbar}\delta(\hbar\omega_{r}-\hbar\Omega)\left\Vert \sum_{m}\mathcal{Q}\left(m,r\right)\left\vert \Psi_{P}\right\rangle \right\Vert ^{2},
\end{equation}
and see that the initial radiation rate depends on how $\left\Vert \sum_{m}\mathcal{Q}\left(m,r\right)\left\vert \Psi_{P}\right\rangle \right\Vert$ scales with increasing $N$. If it scales as $N$ -- that is, there is constructive interference between all the kets $\mathcal{Q}\left(m,r\right)\left\vert \Psi_{\text{P}}\right\rangle$ for different $m$ -- then the initial radiation rate scales as $N^{2}$.

\subsection{Atoms}

We first recall how these results apply to the usual problem of Dicke superradiance, where the pump system consists of $N$ identical two-level atoms and the output system consists of modes of the radiation field. Here we have 
\begin{align}
H_{\text{P}}= & \hbar\bar{\omega}\sum_{m}\sigma_{3}^{m},\nonumber \\
H_{\text{R}}= & \sum_{q}\hbar\omega_{q}b_{q}^{\dagger}b_{q},
\end{align}
where $b_{q}$ is a destruction operator for a mode of the radiation field, and we take $q$ to label both the wave vector $\mathbf{k}$ and the polarization type; $\hbar\bar{\omega}$ is the transition energy of the atoms, and $\sigma_{3}^{m}=\left(\left\vert e\right\rangle _{m}\left\langle e\right\vert _{m}-\left\vert g\right\rangle _{m} \left\langle g\right\vert _{m}\right)/2$, where $\left\vert e\right\rangle _{m}$ and $\left\vert g\right\rangle _{m}$ are respectively the excited and ground states of atom $m$. The interaction Hamiltonian in the dipole approximation is given by
\begin{equation}
V=-\sum_{m}\bm{\mathfrak{D}}^{m}\cdot\mathbf{E(r}_{m}),
\end{equation}
where $\bm{\mathfrak{D}}^{m}=\bm{\nu}_{m}\mathcal{\sigma}_{-}^{m}+\text{H.c.}$ is the dipole moment operator of the $m^{\text{th}}$ atom and $\mathbf{E(r}_{m})$ is the electric field operator at the position $\mathbf{r}_{m}$ of the $m^{\text{th}}$ atom. Here $\bm{\nu}_{m}$ is the vector electric dipole matrix element for the transition, $\mathcal{\sigma}_{-}^{m}=\left\vert g\right\rangle _{m}\left\langle e\right\vert _{m}$ is the lowering operator for the $m^{\text{th}}$ atom, and $\mathbf{E(r)=}\sum_{q}i\mathcal{E}_{q}\bm{\epsilon}_{q}b_{q}e^{i\mathbf{k\cdot r}}+\text{H.c.}$ This operator is defined by the polarization of a plane wave mode $\bm{\epsilon}_{q}$ and the electric field per photon $\mathcal{E}_{q}=\sqrt{\hbar ck_{q}/2\epsilon_{0}\mathcal{V}}$, where $\mathcal{V}$ is the normalization volume. For the energy conserving part of the interaction we find $\mathcal{P}_{m}=\mathcal{\sigma}_{-}^{m}$, $\mathcal{R}_{q}^{\dagger}=b_{q}^{\dagger}$, and $c_{mq}=i\mathcal{E}_{q}\bm{\nu}_{m}\cdot\bm{\epsilon}_{q}e^{i\mathbf{k\cdot r}_{m}}$. To calculate the probability per unit time that a photon is emitted into the state $\left\vert \psi_{r}\right\rangle = b_{r}^{\dagger}\left\vert \text{vac}\right\rangle \equiv \left\vert r\right\rangle$, we identify $\Omega=\bar{\omega}$, and find
\begin{equation}
\frac{\mathfrak{P}_{r}(t)}{t}=\frac{2\pi}{\hbar}\delta(\hbar\omega_{r}-\hbar\bar{\omega})\left\Vert \sum_{m}\mathcal{Q}\left(m,r\right)\left\vert \Psi_{\text{P}}\right\rangle \right\Vert ^{2},\label{FGRresult}
\end{equation}
from \eqref{FGR}, with 
\begin{equation}
\mathcal{Q}\left(m,r\right)=c_{mr}\mathcal{\sigma}_{-}^{m}.\label{Dicke}
\end{equation}
For Dicke's example \cite{dicke54} of $N$ identical atoms localized within a wavelength of light, the $c_{mr}$ are independent of $m$, $c_{mr}\rightarrow c_{r}$, and the sum over $m$ of the $\mathcal{Q}\left(m,r\right)$ is proportional to the sum over $m$ of the $\sigma_{-}^{m}$. The states $\left\vert \Psi_{\text{P}}\right\rangle$ that lead to superradiance are the states $\left\vert JM\right\rangle $, which are completely symmetric with respect to exchange of the atoms, and where $J+M$ atoms are in the excited state, and $J-M$ atoms in the ground state. For these states we find
\begin{equation}
\frac{\mathfrak{P}_{r}(t)}{t}=\frac{2\pi}{\hbar}\delta(\hbar\omega_{r}-\hbar\bar{\omega})\left\vert c_{r}\right\vert ^{2}(J+M)(J-M+1).
\end{equation}
This takes its largest value when $J=N/2$ and $M=0$, corresponding to half of the atoms in their excited states, and scales there as $N^{2}$.

\begin{figure}[hbt]
\includegraphics[width=0.45\textwidth]{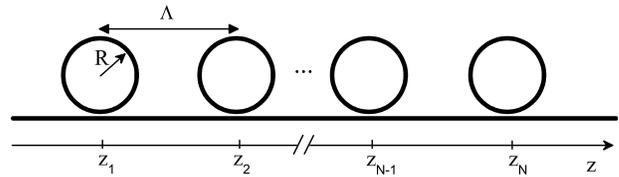}
\caption{The SCISSOR structure.  All $N$ rings have radius $R$ and are equally spaced from their neighbors by a distance $\Lambda$. We position the first ring at $z_{1}$, the second at $z_{2}=z_{1}+\Lambda$, etc.; thus $z_{N}-z_{1}=\left(N-1\right)\Lambda$.}
\label{fig:SCISSOR}
\end{figure}

\subsection{Rings \label{sec:rings}}

\begin{figure}[hbt]
\includegraphics[width=0.45\textwidth]{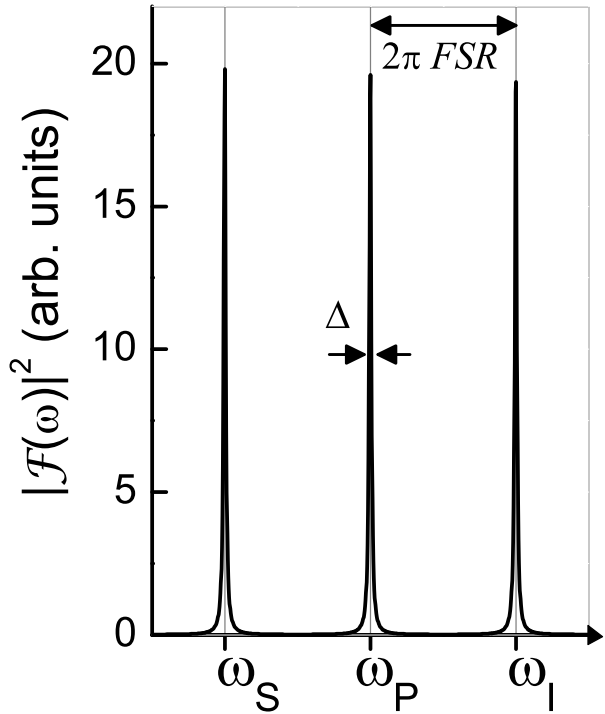}
\caption{Absolute value squared of the field enhancement factor in the ring \eqref{simplerfieldenhancement}, with $\Delta$ the FWHM of the resonances and FSR the free spectral range.}
\label{fig:resonances}
\end{figure}

Next we turn to the SCISSOR structure composed of $N$ rings coupled to a single bus waveguide. For simplicity, we assume that all of the rings are identical, with an equal distance $\Lambda$ between each ring and the following ring, until the $N^{\text{th}}$ ring is reached (see Fig.~\ref{fig:SCISSOR}). To look at super-SFWM in this configuration we take $\mathcal{H}_{\text{P}}$ to be the Hilbert space of the radiation field at pump frequencies, and $\mathcal{H}_{\text{R}}$ the Hilbert space of the radiation field at the signal and idler frequencies. Here $\mathcal{H}_{\text{R}}$ involves only part of the full radiation field, but we refer to that part as the ``radiation subsystem.'' In SFWM the pump frequencies are typically in the neighborhood of one of the resonances $\omega_{\text{P}}$ of the ring resonators, and the signal and idler frequencies are in the neighborhood of resonances at higher and lower frequencies respectively, $\omega_{\text{I}}$ and $\omega_{\text{S}}$ (see Fig.~\ref{fig:resonances}). Light at all these frequencies can in reality pass between the bus waveguide and the rings. However, for our calculation of super-SFWM in this Section we neglect any coupling of light between the rings and the bus waveguide at the \textit{pump} frequencies. Then as an initial state one can imagine pump light ``loaded'' into the rings, and no light present at signal and idler frequencies; at the moment we do not concern ourselves with how this initial state of the rings was prepared, just as in the elementary calculation above of Dicke's system we did not consider the preparation of the initial state of the atoms. In the SCISSOR example our initial state can then also be taken as \eqref{initial}, where now $\left\vert \text{vac}\right\rangle $ refers to the vacuum state of the radiation field at signal and idler frequencies, and $\left\vert \Psi_{\text{P}}\right\rangle $ identifies the initial state of the pump light in the rings. In this idealization, were there no SFWM the pump light would remain in the rings forever, just as if there were no spontaneous emission the atoms introduced by Dicke would remain in their initial state forever. We assume that the resonance modes of the rings of interest are those where light is propagating counterclockwise within the rings.

\begin{figure}[hbt]
\includegraphics[width=0.45\textwidth]{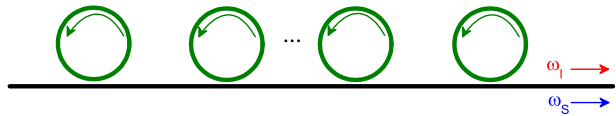}
\caption{A sketch of the Dicke picture of super-SFWM in a SCISSOR, with with strong pump light shown confined to the rings and asymptotic-out signal and idler states.}
\label{fig:DickeSSFWM}
\end{figure}

Within this framework the Hamiltonian for the pump system describes the resonator modes in the rings at the resonance where we consider the (assumed isolated) resonators are ``loaded'' with pump light,
\begin{equation}
H_{\text{P}}=\hbar\omega_{\text{P}}\sum_{m}a_{m}^{\dagger}a_{m},
\end{equation}
where henceforth $\omega_{\text{P}}$ is the pump resonance of interest and $\left[a_{m},a_{m^{\prime}}^{\dagger}\right]=\delta_{mm^{\prime}}$. Quantizing in terms of the electric displacement and magnetic fields \cite{liscidini12}, the displacement field operator in the rings at the pump frequency can then be written as 
\begin{equation}
\mathbf{D}_{\text{P}}(\mathbf{r})=\sum_{m}\sqrt{\frac{\hbar\omega_{\text{P}}}{2}}a_{m}\mathbf{D}_{\text{iso}}(\mathbf{r},m)+\text{H.c.},\label{DP}
\end{equation}
where $\mathbf{D}_{\text{iso}}(\mathbf{r},m)$ is a mode field that exists only in ring $m$, and describes the counterclockwise propagation of light (see Fig.~\ref{fig:DickeSSFWM}).

Through SFWM this will generate light at signal and idler frequencies that will ultimately leave the rings and propagate towards $z=\infty$ in the bus waveguide. To describe this radiation, which here is the ``output system,'' we introduce a normalization length $\mathcal{L}$ of the channel, much longer than the size (about $\left(N-1\right)\Lambda$) of the physical system; we quantize the output system over the length $\mathcal{L}$ under the usual assumption of periodic boundary conditions, and will ultimately take $\mathcal{L}\rightarrow\infty$. The Hamiltonian for the output light is
\begin{equation}
H_{\text{R}}=\sum_{k\in\left\{ k_{s}\right\} }\hbar\omega(k)b_{k}^{\dagger}b_{k}+\sum_{k\in\left\{ k_{i}\right\} }\hbar\omega(k)b_{k}^{\dagger}b_{k},
\end{equation}
where $\left[b_{k},b_{k^{\prime}}^{\dagger}\right]=\delta_{kk^{\prime}}$, with $\left\{ k_{s}\right\} $ and $\left\{ k_{i}\right\}$ respectively the sets of signal and idler frequencies under the appropriate resonances. The wave number $k$ here labels the different modes, with frequencies $\omega(k)$. Since the signal and idler light propagates through the rings as well as in the bus waveguide, and since we are interested in signal and idler light being generated and leaving the system, we take these modes as ``asymptotic-out'' states of the radiation field \cite{liscidini12}, familiar from scattering theory \cite{breit54}. In this example, where we are interested in calculating signal and idler light propagating towards $z=+\infty$, the asymptotic-out states are chosen so that for $z$ greater than the position of any of the rings but within the central normalization length the fields are propagating with wave number $k$ and normalized as would be the fields propagating in an isolated bus waveguide; their behavior for smaller $z$ along the waveguide, and within the rings, then follows from the Maxwell equations. From a standard introduction of these asymptotic-out fields \cite{liscidini12} we write the displacement operator associated with them as
\begin{align}
\mathbf{D}_{\text{O}}(\mathbf{r})= & \sqrt{\frac{2\pi}{\mathcal{L}}}\sum_{k\in k_{s}}\sqrt{\frac{\hbar\omega(k)}{2}}b_{k}\mathbf{D}_{\text{R}k}^{\text{asy-out}}(\mathbf{r})+\text{H.c.}\nonumber \\
 & +\sqrt{\frac{2\pi}{\mathcal{L}}}\sum_{k\in k_{i}}\sqrt{\frac{\hbar\omega(k)}{2}}b_{k}\mathbf{D}_{\text{R}k}^{\text{asy-out}}(\mathbf{r})+\text{H.c.},\label{DO}
\end{align}
(see Fig.~\ref{fig:DickeSSFWM}) where S and I refer to signal and idler photons respectively, and where the subscript R on the asymptotic-out mode $\mathbf{D}_{\text{R}k}^{\text{asy-out}}(\mathbf{r})$ indicates light leaving the system to the right; while such an expansion is usually done where $k$ ranges continuously \cite{liscidini12}, the overall factor $\sqrt{2\pi/\mathcal{L}}$ here accounts that for the moment we are quantizing within a normalization length $\mathcal{L}$.

The nonlinearity responsible for SFWM is characterized by the usual third order susceptibility tensor $\chi_{3}$ \cite{agrawal07}. Writing the nonlinear Hamiltonian in terms of the displacement fields instead \cite{sipe04}, the appropriate $V$ appearing in \eqref{Htotal} is
\begin{equation}
V=-\frac{1}{4\varepsilon_{0}}\sum_{m}\int_{V_{m}}\text{d}\mathbf{r}\,\Gamma_{3}^{ijkl}(\mathbf{r})D^{i}(\mathbf{r})D^{j}(\mathbf{r})D^{k}\left(\mathbf{r}\right)D^{l}\left(\mathbf{r}\right),\label{Vring}
\end{equation}
where $\mathbf{D(r)}$ is the total displacement operator, and we neglect dispersion in the nonlinearity; $\Gamma_{3}^{ijkl}(\mathbf{r})$ is related to $\chi_{3}^{ijkl}(\mathbf{r})$, and the integral over $V_{m}$ indicates an integral over the $m^{\text{th}}$ ring. Using \eqref{DP} and \eqref{DO} in \eqref{Vring}, we keep terms of the form $b_{k_{s}}^{\dagger}b_{k_{i}}^{\dagger}a_{m}a_{m}+\text{H.c.}$, where two pump photons are converted to a signal and idler photon, and find
\begin{align}
V= & -\frac{2\pi}{\mathcal{L}}\sum_{m,k_{s},k_{i}}\frac{3\hbar^{2}\omega_{\text{P}}\sqrt{\omega_{s}\omega_{i}}}{4\varepsilon_{0}}\nonumber \\
 & \times F^{(m)}(\omega_{s},\omega_{i})b_{k_{s}}^{\dagger}b_{k_{i}}^{\dagger}a_{m}a_{m}+\text{H.c.},\label{Vringsuse}
\end{align}
where $k_{s(i)}$ ranges over the wave numbers associated with signal (idler) light, with $\omega_{s(i)}\equiv\omega(k_{s(i)})$ the associated frequencies, and
\begin{align}
F^{(m)}(\omega_{s},\omega_{i})= & \int_{V_{m}}\text{d}\mathbf{r}\,\Gamma_{3}^{ijkl}(\mathbf{r})\left[D_{\text{R}k_{s}}^{i,\text{asy-out}}(\mathbf{r})\right]^{\ast}\nonumber \\
 & \times\left[D_{\text{R}k_{i}}^{j,\text{asy-out}}(\mathbf{r})\right]^{\ast}D_{\text{iso}}^{k}(\mathbf{r},m)D_{\text{iso}}^{l}(\mathbf{r},m).\label{Fdef}
\end{align}

The interaction Hamiltonian \eqref{Vringsuse} is of the form introduced above \eqref{Vform}, where the label $q$ is now identified with the pair $(k_{s},k_{i})$, the wave numbers of the signal and idler respectively, $\mathcal{P}_{m}=a_{m}a_{m}$, $\mathcal{R}_{(k_{s},k_{i})}^{\dagger}=b_{k_{s}}^{\dagger}b_{k_{i}}^{\dagger}$, and
\begin{equation}
c_{m(k_{s},k_{i})}=\frac{2\pi}{\mathcal{L}}\frac{3\hbar^{2}\omega_{\text{P}}\sqrt{\omega_{s}\omega_{i}}}{4\varepsilon_{0}}F^{(m)}(\omega_{s},\omega_{i}).
\label{eq:cmk}
\end{equation}
Here $\Omega=2\omega_{\text{P}}$, and the probability per unit time that a signal and idler photon are emitted to form the state of a pair of photons $\left\vert \psi_{r}\right\rangle =b_{k_{s}}^{\dagger}b_{k_{i}}^{\dagger}\left\vert \text{vac}\right\rangle $, with energy $\hbar\omega_{r}\equiv\hbar\omega_{s}+\hbar\omega_{i}$, is given by \eqref{FGR} as
\begin{equation}
\frac{\mathfrak{P}_{r}(t)}{t}=\frac{2\pi}{\hbar}\delta(\hbar\omega_{r}-2\hbar\omega_{\text{P}})\left\Vert \sum_{m}\mathcal{Q}\left(m,r\right)\left\vert \Psi_{\text{P}}\right\rangle \right\Vert ^{2},\label{ringresult}
\end{equation}
with $\mathcal{Q}\left(m,r\right)=c_{m(k_{s},k_{i})}a_{m}a_{m}$.

We now see how super-SFWM can arise, as well as its analogy with the Dicke superradiance of a collection of atoms. Suppose that the light at the pump frequency in each ring is a coherent state, with $\left\vert \Psi_{\text{P}}\right\rangle =\left\vert \left\{ \alpha_{j}\right\} \right\rangle $, where $\left\{ \alpha_{j}\right\} $ denotes the set $\left\{ \alpha_{1},\alpha_{2},\ldots,\alpha_{N}\right\} $, with
\begin{equation}
a_{m}\left\vert \left\{ \alpha_{j}\right\} \right\rangle =\alpha_{m}\left\vert \left\{ \alpha_{j}\right\} \right\rangle .\label{alpham}
\end{equation}
Then
\begin{equation}
\sum_{m}\mathcal{Q}\left(m,r\right)\left\vert \Psi_{\text{P}}\right\rangle =\left(\sum_{m}c_{m(k_{s},k_{i})}\alpha_{m}^{2}\right)\left\vert \Psi_{\text{P}}\right\rangle .\label{amwork}
\end{equation}
While in Dicke's example the atoms were all assumed to be identical and within a distance of each other that is much less than the wavelength of light, and thus in the corresponding $\mathcal{Q}\left(m,r\right)$ \eqref{Dicke} the $c_{mr}$ could be taken to be independent of $m$, here even if the rings are all identical the distance from one ring to any of the others will be much larger than the wavelength of light, simply because each ring itself is larger than the wavelength of light. Hence the $F^{(m)}(\omega_{s},\omega_{i})$ will in general be different for each $m$, even though we have assumed the rings are identical, for the asymptotic-out field $\mathbf{D}_{\text{R}k_{s(i)}}^{\text{asy-out}}(\mathbf{r})$ in one ring will be different compared to that in another due to a phase difference arising from propagation between rings in the bus waveguide. Thus the $c_{m(k_{s},k_{i})}$ \eqref{eq:cmk} for different $m$ will in general differ by a phase. Nonetheless, one can imagine that if the phases of the $\alpha_{m}$ are adjusted to compensate for this, and the $\left\vert \alpha_{m}\right\vert $ is the same for each $m$, then
\begin{equation}
\sum_{m}c_{m(k_{s},k_{i})}\alpha_{m}^{2}\propto N,\label{ringsum}
\end{equation}
where $N$ is the number of rings, and super-SFWM will arise, with a simple description very similar to that of Dicke superradiance. Identifying the conditions under which this will occur, and constructing a more realistic description of the state of the pump light in the rings, is the goal of the following Sections.

\section{The long pump pulse limit \label{sec:longpulse}}

A natural limit to look for agreement between the previous section and a more sophisticated treatment of nonlinear quantum optics in a SCISSOR is that of a long pump pulse.  This is because a Fermi's Golden Rule calculation results in a transition {\em{rate}}, and as the excitation becomes effectively CW the probability of pair generation per pump pulse can also be written as a rate.  Thus this is where we begin a more detailed investigation of super-SFWM.  

\textcolor{red}{In general, a SFWM process in the presence of a pump pulse leads to a multimode squeezed state.  However, in the limit of a small squeezing parameter,  $\left\vert\beta\right\vert$, this state can be approximated as
\begin{equation}
\left\vert\psi\right\rangle\approx\left\vert\text{vac}\right\rangle+\frac{\beta}{\sqrt{2}}\int\text{d}\omega_{1}\text{d}\omega_{2}\,\phi\left(\omega_{1},\omega_{2}\right)a_{\omega_{1}}^{\dagger}a_{\omega_{2}}^{\dagger}\left\vert\text{vac}\right\rangle,
\end{equation}
where $\phi\left(\omega_{1},\omega_{2}\right)$ is the normalized biphoton wave function (BWF), allowing interpretation of $\left\vert\beta\right\vert^{2}$ as the probability of generating a pair of photons per pump pulse. Neglecting time-ordering corrections~\cite{Quesada2015} and other nonlinear effects we note that this BWF provides complete knowledge of the full state of generated photons, and thus enables calculation of arbitrary field expectation values and correlation functions~\cite{christ11}.}

As we have previously sketched \cite{helt12_2}, and detail in Appendix \ref{sec:formalism} of this work, the normalized BWF for a SCISSOR of $N$ rings can be written as
\begin{align}
\phi(\omega_{1},\omega_{2})= & \frac{3\pi i\sqrt{2}\alpha^{2}\hbar}{4\varepsilon_{0}\beta}\sqrt{\frac{\omega_{1}\omega_{2}}{v(\omega_{1})v(\omega_{2})}}\nonumber \\
 & \times\int\text{d}\omega\,\sqrt{\frac{\omega(\omega_{1}+\omega_{2}-\omega)}{v(\omega)v(\omega_{1}+\omega_{2}-\omega)}}\phi_{\text{P}}(\omega)\nonumber \\
 & \times\phi_{\text{P}}(\omega_{1}+\omega_{2}-\omega)J(\omega_{1},\omega_{2},\omega,\omega_{1}+\omega_{2}-\omega),\label{biphoton}
\end{align}
where $\phi_{\text{P}}(\omega)$ is the pump pulse waveform and is normalized according to
\begin{equation}
\int\text{d}\omega\left\vert\phi_{\text{P}}\left(\omega\right)\right\vert^2=1. \label{eq:phiPnorm}
\end{equation}
Here $J\left(\omega_{1},\omega_{2},\omega_{3},\omega_{4}\right)$ is a generalized phase matching function defined later, the group velocities $v\left(\omega_{n}\right)=\left.\text{d}\omega/\text{d}k\right|_{\omega=\omega_{n}}$, and $\left|\alpha\right|^{2}$ is the average number of photons per pump pulse. The probability $\left\vert \beta\right\vert ^{2}$ that a pair of photons is generated can be extracted from the normalization condition
\begin{equation}
\int\left\vert \phi\left(\omega_{1},\omega_{2}\right)\right\vert ^{2}\text{d}\omega_{1}\text{d}\omega_{2}=1,\label{norm}
\end{equation}
and is given explicitly by
\begin{align}
\left\vert \beta\right\vert ^{2}= & \left\vert \frac{3\pi\sqrt{2}\alpha^{2}\hbar}{4\varepsilon_{0}}\right\vert ^{2}\int\text{d}\omega_{1}\text{d}\omega_{2}\frac{\omega_{1}\omega_{2}}{v(\omega_{1})v(\omega_{2})}\nonumber \\
 & \times\left\vert \int\text{d}\omega\,\sqrt{\frac{\omega(\omega_{1}+\omega_{2}-\omega)}{v(\omega)v(\omega_{1}+\omega_{2}-\omega)}}\phi_{\text{P}}(\omega)\right.\nonumber \\
 & \times\left.\phi_{\text{P}}(\omega_{1}+\omega_{2}-\omega)J(\omega_{1},\omega_{2},\omega,\omega_{1}+\omega_{2}-\omega)\vphantom{\sqrt{\frac{\omega(\omega_{1}+\omega_{2}-\omega)}{v(\omega)v(\omega_{1}+\omega_{2}-\omega)}}}\right\vert ^{2}.\label{betasquaredgeneral}
\end{align}
The long pump pulse limit of these expressions allows a strong connection with the Fermi's Golden Rule result of the previous Section. 

We first consider a pump pulse of length $\Delta T$ idealized as a ``top hat function'' with a center frequency $\omega_{\text{P}}$,
\begin{align}
\phi_{\text{P}}(t)= & \frac{e^{-i\omega_{\text{P}}t}}{\sqrt{\Delta T}}\text{ for }-\frac{\Delta T}{2}<t<\frac{\Delta T}{2},\nonumber \\
= & 0\text{ otherwise.}\label{tophat}
\end{align}
We take $\omega_{\text{P}}$ to be one of the ring resonances. The spectral amplitude $\phi_{\text{P}}(\omega)$ is then given by
\begin{align}
\phi_{\text{P}}(\omega)= & \int\frac{\text{d}t}{\sqrt{2\pi}}\phi_{\text{P}}(t)e^{i\omega t}\nonumber \\
= & \frac{1}{\sqrt{\pi\Delta\omega}}\text{sinc}\left(\frac{\omega-\omega_{P}}{\Delta\omega}\right),\label{phipulse}
\end{align}
where $\Delta\omega\equiv2/\Delta T$ and sinc$(x)=(\sin x)/x$, satisfying \eqref{eq:phiPnorm}. In Appendix \ref{sec:longpulsedetails} we show that for \textcolor{red}{$\Delta\omega$ sufficiently narrow -- i.e. $\Delta T$ long enough compared to other time scales in the problem, in particular compared to the dwelling time of a photon in a ring -- the expression \eqref{betasquaredgeneral} is well-approximated by}
\begin{align}
\frac{\left\vert \beta\right\vert ^{2}}{\left\vert \alpha\right\vert ^{4}}\approx & \frac{1}{\Delta T}\frac{9\pi^{3}}{2\varepsilon_{0}^{2}}\left(\frac{\hbar\omega_{\text{P}}}{v(\omega_{\text{P}})}\right)^{2}\nonumber \\
 & \times\int\text{d}\omega\,\frac{\omega(2\omega_{\text{P}}-\omega)}{v(\omega)v(2\omega_{\text{P}}-\omega)}\left\vert J(\omega,2\omega_{\text{P}}-\omega,\omega_{\text{P}},\omega_{\text{P}})\right\vert ^{2},\label{betasquared1}
\end{align}
where the integration over $\omega$ now ranges only over the frequency range of the signal (see Fig. \ref{fig:resonances}). For a fixed number of rings the only dependence of the right-hand-side on the pump pulse is through its dependence on $(\Delta T)^{-1}$. Thus, for fixed $\left\vert \alpha\right\vert ^{2}$, the probability that a pair of down-converted photons is generated is inversely proportional to the length of the pulse. This scaling arises simply because $\hbar\omega_{\text{P}}\left\vert \alpha\right\vert ^{2}$ is approximately the energy in the long pump pulse; the energy \textit{density} $\mathfrak{E}$ (energy per unit length) of the incident pump pulse is
\begin{equation}
\mathfrak{E}=\frac{\hbar\omega_{\text{P}}\left\vert \alpha\right\vert ^{2}}{v(\omega_{\text{P}})\,\Delta T},
\end{equation}
and we can write \eqref{betasquared1} as
\begin{align}
\frac{\left\vert \beta\right\vert ^{2}}{\Delta T}\approx & \frac{9\pi^{3}}{2\varepsilon_{0}^{2}}\mathfrak{E}^{2}\nonumber \\
 & \times\int\text{d}\omega\frac{\omega(2\omega_{P}-\omega)}{v(\omega)v(2\omega_{P}-\omega)}\left\vert J(\omega,2\omega_{\text{P}}-\omega,\omega_{\text{P}},\omega_{\text{P}})\right\vert ^{2},\label{betasquared2}
\end{align}
showing that the \textit{rate} of photon pair production depends on the square of the energy density of the pump pulse, as expected for spontaneous four-wave mixing. At the same level of approximation we find in Appendix \ref{sec:longpulsedetails} that the expression for the joint spectral density is
\begin{align}
\left\vert \phi\left(\omega_{1},\omega_{2}\right)\right\vert ^{2}\approx & \mathcal{N}\frac{\omega_{1}\omega_{2}}{v(\omega_{1})v(\omega_{2})}\nonumber \\
 & \times\left\vert J\left(\omega_{1},\omega_{2},\omega_{\text{P}},\omega_{\text{P}}\right)\right\vert ^{2}\delta(2\omega_{\text{P}}-\omega_{1}-\omega_{2}),\label{JSDlong}
\end{align}
where $\omega_{1}$ and $\omega_{2}$ can range over both signal and idler frequencies, and $\mathcal{N}$ is a constant to guarantee the normalization condition \eqref{norm}.

The presence of the energy-conserving Dirac delta function in \eqref{JSDlong}, and the prediction \eqref{betasquared2} of a rate of photon pair emission, suggest that the long pump pulse limit identified here should be described by the simple Fermi's Golden Rule argument given in Section \ref{sec:rings}. However, to make that connection we have to relate the pair production rate \eqref{betasquared2} not to the square of the energy density in the pump pulse, but to the square of the energy $\hbar\omega_{\text{P}}\left\vert \alpha_{m}\right\vert ^{2}$ that the pump pulse ``loads'' into each ring (recall \eqref{DP},\eqref{alpham}), since it is these $\alpha_{m}$ that appear in \eqref{ringresult}, \eqref{amwork}. We do this in Appendix \ref{sec:connection}, and indeed find that \eqref{betasquared2} is equivalent to the result \eqref{ringresult} for the Fermi's Golden Rule calculation.

What is yet to be established is how the rate of photon pair emission scales with the number of rings. We address that issue in the next Section, for both the long pump pulse limit that we have shown is described by the simple Fermi's Golden Rule calculation, and for shorter pump pulses where it is necessary to use the more general formalism.

\section{The coherence of the rings \label{sec:coherence}}

To evaluate the pair production rate, either in the long pump pulse limit \eqref{betasquared2} or more generally \eqref{betasquaredgeneral}, we now examine the generalized phase matching function
\begin{align}
J(\omega_{1},\omega_{2},\omega_{3},\omega_{4})= & e^{i\chi}\frac{\sin\frac{\mu N}{2}}{\sin\frac{\mu}{2}}\nonumber \\
 & \times j^{\text{ref}}\left[k(\omega_{1}),k(\omega_{2}),k(\omega_{3}),k(\omega_{4})\right], \label{JRRLLdefmaintext}
\end{align}
which is derived in Appendix \ref{sec:formalism}; the phases $\chi\left(\omega_{1},\omega_{2},\omega_{3},\omega_{4}\right)$ and $\mu\left(\omega_{1},\omega_{2},\omega_{3},\omega_{4}\right)$ are discussed below. The two crucial components of \eqref{JRRLLdefmaintext} are $j^{\text{ref}}\left[k(\omega_{1}),k(\omega_{2}),k(\omega_{3}),k(\omega_{4})\right]$, the contribution from a single ring, and the factor involving $\mu(\omega_{1},\omega_{2},\omega_{3},\omega_{4})$ that identifies to what extent the contributions from the different rings add in phase. If $\mu=0$ then there is complete constructive interference, or $\left\vert J\right\vert^2=N^2\left\vert j^{\text{ref}}\right\vert^2$, resulting in an $N^{2}$ scaling of the pair generation rate. We begin by simplifying the expression for $j^{\text{ref}}\left[k(\omega_{1}),k(\omega_{2}),k(\omega_{3}),k(\omega_{4})\right]$, since the resonance structure of this factor in $J(\omega_{1},\omega_{2},\omega_{3},\omega_{4})$ identifies approximations that can be made in evaluating the other factors.

\begin{figure}[hbt]
\includegraphics[width=0.45\textwidth]{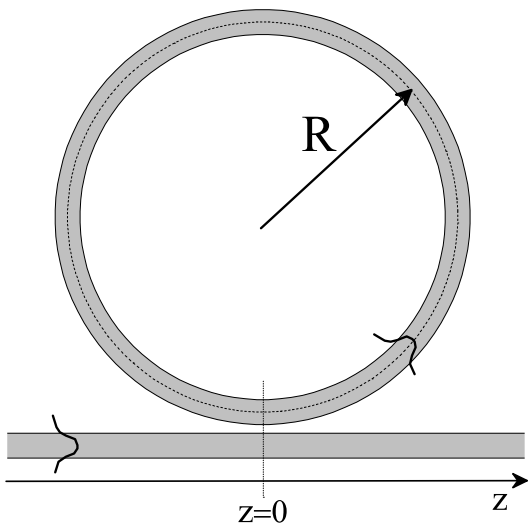}
\caption{A single ring of radius $R$ positioned at $z=0$ with transverse field profiles in the ring and channel indicated by Gaussians.}
\label{fig:singleSCISSOR}
\end{figure}

As detailed in Appendix \ref{sec:formalism}, we can write the contribution from a single ring as
\begin{align}
 & j^{\text{ref}}\left[k(\omega_{1}),k(\omega_{2}),k(\omega_{3}),k(\omega_{4})\right]\nonumber \\
 & =\int \text{d}\mathbf{r}\Gamma_{(3)}^{ijkl}(\mathbf{r})\left[D_{k\left(\omega_1\right)}^{i}\left(\mathbf{r}\right)D_{k\left(\omega_2\right)}^{j}\left(\mathbf{r}\right)\right]^{\ast}\nonumber \\ &\quad\times D_{k\left(\omega_3\right)}^{k}\left(\mathbf{r}\right)D_{k\left(\omega_4\right)}^{l}\left(\mathbf{r}\right), \label{jwork}
\end{align}
where the components of the displacement fields 
\begin{equation}
D_{k\left(\omega\right)}^i\left(\mathbf{r}\right)=\mathfrak{F}\left(\omega\right)\frac{\mathcal{D}^i\left(\mathbf{r}\right)}{\sqrt{2\pi}}e^{i k(\omega)\zeta},
\end{equation}
are those inside the ring and $\zeta$ ranges from $0$ to $l$. Here the 
\begin{equation}
\mathfrak{F}(\omega)=\frac{i\kappa}{1-\sigma e^{ik(\omega)l}},\label{simplerfieldenhancement}
\end{equation}
are field enhancement factors with $\kappa$ and $\sigma$ the usual cross- and self-coupling coefficients \cite{heebner08}, and $l=2\pi R$ the circumference of the ring with radius $R$. We neglect the weak wave number dependence of the $\mathcal{D}(\mathbf{r})$ in \eqref{jwork}; more general expressions are given in Appendix \ref{sec:formalism}. The integral in \eqref{jwork} is over a ring imagined just above the bus waveguide at $z=0$ (see Fig.~\ref{fig:singleSCISSOR}). Similarly, we neglect any weak dependence of $\sigma$ and $\kappa$ on $k(\omega)$ in the field enhancement factors $\mathfrak{F}\left(\omega\right)$ and consider the physically relevant case of weak coupling when
\begin{align}
\left\vert \kappa\right\vert \ll & \left\vert \sigma\right\vert ,\nonumber \\
\left\vert \sigma\right\vert \lesssim & 1.\label{WC}
\end{align}
We consider both $\kappa$ and $\sigma$ positive, though note that generalizing this is straightforward. Resonances occur at frequencies $\omega_{M}$ where
\begin{equation}
k(\omega_{M})l=2\pi M\text{,}\label{rfrequencies}
\end{equation}
with $M$ an integer, and for $\omega$ close to $\omega_{M}$ the field enhancement factor \eqref{simplerfieldenhancement} simplifies to
\begin{equation}
\mathfrak{F}\left(\omega\right)\rightarrow\sqrt{\frac{2}{1-\sigma}}\frac{\Delta/2}{(\omega-\omega_{M})-i(\Delta/2)},
\end{equation}
where
\begin{equation}
\Delta=\frac{2(1-\sigma)v_{g}}{l},\label{FWHM}
\end{equation}
is the FWHM of the resonance in $\left\vert \mathfrak{F}\left(\omega\right)\right\vert ^{2}$ at $\omega_{M}$. We sketch $\left\vert \mathfrak{F}\left(\omega\right)\right\vert ^{2}$ near our resonances of interest in Fig.~\ref{fig:resonances}. In arriving at this expression, we have assumed that $\omega$ is sufficiently close to $\omega_{M}$ that group velocity dispersion can be neglected, with the group velocity taken to be $v_{g}$. Indeed, we make the stronger assumption that over the frequency range including the pump, signal, and idler resonances we can neglect group velocity dispersion, writing $v(\omega)=v_{g}$ for all such frequencies, where
\begin{equation}
k(\omega)=k(\omega_{M})+\frac{\omega-\omega_{M}}{v_{g}},\label{noGVD}
\end{equation}
with $\omega_{M}$ any resonant frequency in this range. This approximation is justified for realistic structures, as detailed below \cite{helt12_2}.  With this approximation  the free spectral range is
\begin{equation}
FSR=\frac{\omega_{M+1}-\omega_{M}}{2\pi}=\frac{v_{g}}{l}.\label{FSR}
\end{equation}
In the weak coupling limit \eqref{WC} it is also true that
\begin{equation}
\frac{\Delta}{2\pi}<<FSR
\end{equation}
(see Fig.~\ref{fig:resonances}). Taking a signal ring resonance frequency $\omega_{\text{S}}$ and an idler ring resonance frequency $\omega_{\text{I}}$ of interest, with frequencies smaller and larger than the pump ring resonance frequency $\omega_{\text{P}}$, respectively, such that $2k(\omega_{\text{P}})-k(\omega_{\text{S}})-k(\omega_{\text{I}})=0$, \eqref{noGVD} tells us that
\begin{equation}
2\omega_{\text{P}}-\omega_{\text{S}}-\omega_{\text{I}}=0,\label{rescond}
\end{equation}
for the resonance frequencies themselves. For $\omega_{1}$ close to $\omega_{S}$, $\omega_{2}$ close to $\omega_{I}$, and both $\omega_{3}$ and $\omega_{4}$ close to $\omega_{\text{P}}$, \eqref{jwork} reduces to
\begin{align}
 & j^{\text{ref}}\left[k(\omega_{1}),k(\omega_{2}),k(\omega_{3}),k(\omega_{4})\right]\nonumber \\
 & =\left(\frac{2}{1-\sigma}\right)^{2}\frac{\Delta/2}{(\omega_{1}-\omega_{\text{S}})-i(\Delta/2)}\frac{\Delta/2}{(\omega_{2}-\omega_{\text{I}})-i(\Delta/2)}\nonumber \\
 & \quad\times\frac{\Delta/2}{(\omega_{3}-\omega_{\text{P}})-i(\Delta/2)}\frac{\Delta/2}{(\omega_{4}-\omega_{\text{P}})-i(\Delta/2)}\nonumber \\
 & \quad\times\int\frac{\text{d}\mathbf{r}}{(2\pi)^{2}}\Gamma_{(3)}^{ijkl}(\mathbf{r})\left[\mathcal{D}^{i}(\mathbf{r})\mathcal{D}^{j}(\mathbf{r})\right]^{\ast}\mathcal{D}^{k}(\mathbf{r})\mathcal{D}^{l}(\mathbf{r})\nonumber \\
& \quad\times e^{i(\omega_{3}+\omega_{4}-\omega_{1}-\omega_{2})\zeta v_{g}^{-1}}.\label{jrefwork}
\end{align}
We use this expression below to evaluate $J(\omega_{1},\omega_{2},\omega_{3},\omega_{4})$ for various pump pulse durations.

Within the simple Fermi's Golden Rule description of Section \ref{sec:superradiance} the interference between the biphoton amplitudes generated for each ring depends on the $c_{m}(k_{s},k_{i})\alpha_{m}^{2}$ of \eqref{ringsum}, and whether or not they add constructively depends on the relative phase of the different amplitudes $\alpha_{m}$ describing the excitation at the pump frequency in the rings. In the long pump pulse limit, the calculation of Section \ref{sec:longpulse} identifies how the rings are actually ``loaded'' with pump light, and using \eqref{betasquared2} we can examine what kind of radiance is actually exhibited for such loading. The required form of the generalized phase matching function $J(\omega,2\omega_{\text{P}}-\omega,\omega_{\text{P}},\omega_{\text{P}})$ can be extracted as a special case of the more general problem considered below, but we show in Appendix \ref{sec:connection} explicitly that under our assumptions of wave number independent coupling coefficients and weak coupling \eqref{WC}, at the frequencies required $\mu$ in fact vanishes and thus $J(\omega,2\omega_{\text{P}}-\omega,\omega_{\text{P}},\omega_{\text{P}})$ is proportional to $N$. So a long enough pump pulse will indeed ``load'' the rings in such a way that the spontaneous four-wave mixing will indeed be superradiant with a generation rate of photon pairs scaling as $N^{2}$; the final result to be used in \eqref{betasquared2} is
\begin{align}
&\left\vert J(\omega,2\omega_{\text{P}}-\omega,\omega_{\text{P}},\omega_{\text{P}})\right\vert ^{2}\nonumber \\
& = N^{2}\left(\frac{2}{1-\sigma}\right)^{4}\left(\frac{(\Delta/2)^{2}}{(\omega-\omega_{\text{S}})^{2}+(\Delta/2)^{2}}\right)^{2}\nonumber \\
& \quad\times\left\vert \int\frac{\text{d}\mathbf{r}}{(2\pi)^{2}}\Gamma_{(3)}^{ijkl}(\mathbf{r})\left[\mathcal{D}^{i}(\mathbf{r})\mathcal{D}^{j}(\mathbf{r})\right]^{\ast}\mathcal{D}^{k}(\mathbf{r})\mathcal{D}^{l}(\mathbf{r})\right\vert ^{2},\label{J2CWresult}
\end{align}
and the joint spectral density then follows from \eqref{JSDlong}. Of course, for signal and idler sufficiently far in frequency from the pump, group velocity dispersion need be taken into account and the full coherence of the spontaneous emission rate will be lost, but that can be described by a straight-forward generalization of the above equations. We have considered the effect of group velocity dispersion earlier \cite{helt12_2}, and found that for a typical dispersion in silicon structures, $\xi\simeq0.01\text{ ps}^{2}\text{/m}$ {[}$-8$ ps/(nm$\cdot$km){]}, about a thousand rings would be necessary for degradation in the scaling as $N^{2}$ to set in. Thus we are indeed justified in neglecting group velocity dispersion throughout this Section.

More interesting are the generation rate and joint spectral density that will result if we consider shorter pump pulses. For that we return to the more general equations \eqref{biphoton} and \eqref{betasquaredgeneral}, for $\phi(\omega_{1},\omega_{2})$ and $\left\vert \beta\right\vert ^{2}$, respectively. In these expressions $\omega_{1}$ and $\omega_{2}$ can both range over signal and idler photon frequencies, but we focus on the contributions where the first of these is close to a signal ring resonance frequency $\omega_{\text{S}}$ and the second is close to an idler ring resonance frequency $\omega_{\text{I}}$; the term where $\omega_{1}$ and $\omega_{2}$ are interchanged can be considered similarly. Then the quantity of interest is $J(\omega_{1},\omega_{2},\omega_{3},\omega_{4})$ where $\omega_{3}$ and $\omega_{4}$ are pump pulse frequencies, and the energy conservation constraint implicit in the general equations \eqref{biphoton} and \eqref{betasquaredgeneral} identifies
\begin{equation}
\omega_{4}=\omega_{1}+\omega_{2}-\omega_{3}.\label{energyconservation}
\end{equation}
The quantity $J(\omega_{1},\omega_{2},\omega_{3},\omega_{4})$, with $\omega_{4}$ restricted by (\ref{energyconservation}) is a richer function than the function of a single variable $J(\omega,2\omega_{\text{P}}-\omega,\omega_{\text{P}},\omega_{\text{P}})$ required in the long pump pulse limit. 

Rather than considering just the ``top hat function'' pulse of Section \ref{sec:longpulse}, we consider a more general pulse characterized by frequency amplitudes $\phi_{\text{P}}(\omega)$, with significant amplitudes only for $\omega$ within a range $\delta\ll\Delta$ of a pump ring resonance frequency $\omega_{\text{P}}$,
\begin{equation}
\left\vert \omega-\omega_{\text{P}}\right\vert \leq\delta/2,\label{pumpconstraints}
\end{equation}
and phase variation small over that range. This allows us to consider a pump pulse short enough to explore how $\mu$ becomes nonzero as the pump pulse duration departs from the long pulse limit, and yet still long enough to take full advantage of the field enhancement in the ring so that there is a significantly enhanced spontaneous emission rate.  Defining
\begin{equation}
u\equiv(\omega_{1}-\omega_{\text{S}})+(\omega_{2}-\omega_{\text{I}}).\label{upsilondef}
\end{equation}
we use the energy conservation condition \eqref{energyconservation} and the resonance frequency condition \eqref{rescond} to write $u=\omega_{3}+\omega_{4}-2\omega_{\text{P}}$ and then, noting the constraints \eqref{pumpconstraints} that hold for both pump pulse waveforms appearing in \eqref{biphoton} and \eqref{betasquaredgeneral}, we calculate that significant contributions occur only for $\omega_{1}$ and $\omega_{2}$ satisfying $\left\vert u\right\vert \lesssim\delta$.  

\textcolor{red}{In the limit~\eqref{noGVD} of negligible group velocity dispersion, as well as that of weak coupling~\eqref{WC}, we calculate (see Appendix~\ref{sec:coherencenumber}) that the $\mu$ of~\eqref{JRRLLdefmaintext} has a typical size of about
\begin{equation}
\left\vert\mu\right\vert\approx\frac{2\delta}{\Delta}.\label{eq:mumaintext}
\end{equation}}%
Were $\mu$ truly constant, then as the number of rings $N$ were increased we would begin to see the phase-matching function \eqref{JRRLLdefmaintext} decrease when $N=N_{\text{coh}}$, with 
\begin{equation}
N_{\text{coh}}\left\vert \mu\right\vert=\pi,
\end{equation}
or
 \begin{equation}
N_{\text{coh}}\approx\frac{\pi}{2}\frac{\Delta}{\delta}\approx\frac{\pi}{2}\frac{\tau_{\text{pump}}}{\tau_{\text{dwell}}},\label{Ncohresult}
\end{equation}
where $\tau_{\text{pump}}\equiv1/\delta$ is approximately the pump pulse duration, and $\tau_{\text{dwell}}\equiv1/\Delta$ is approximately the dwelling time of light in a ring. Of course $\mu$ is not constant, and for $N$ very large we expect the probability of the generation of a photon pair to scale with $N$, as coherence between the rings vanishes. Nonetheless, we can take $N\approx N_{\text{coh}}$ to be a somewhat heuristic identification of when the dependence of that probability changes from $N^{2}$ at small $N$ to $N^{1}$ at large $N$. 

We close this Section with some calculations of the pair generation rate and joint spectral densities for a typical silicon-on-insulator SCISSOR structure with parameters corresponding to operation in the telecommunication wavelength range \cite{agrawal07}. We consider a structure of $N$ identical rings with radius $R=5\text{ }\mu\text{m}$, separated by a distance $\Lambda=15\text{ }\mu\text{m}$. We take an effective (phase) index $\bar{n}=2.5$ and a group velocity $v_{g}=0.75\times10^{8}\text{ m/s}$, corresponding to a group index $n_{g}=4$; we neglect group velocity dispersion in our calculations. We take the pump resonance to be the $50^{\text{th}}$, which corresponds to a vacuum wavelength of about $1570\text{ nm}$; the signal and idler resonances are taken to be those immediately above and below. The self-coupling $\sigma$ between the channel and the rings is given by $1-\sigma=0.0126$, which leads to a FWHM (\ref{FWHM}) of $\Delta=6\times10^{10}\text{ rad/s}$; for the pump resonance this gives a $Q$ factor of $Q=\omega_{\text{P}}/\Delta\approx20,000$.

In the absence of group velocity dispersion and dispersion in the nonlinear response, the relation between $\Gamma_{3}^{ijkl}(\mathbf{r)}$ and the usual $\chi_{3}^{ijkl}(\mathbf{r})$ is \cite{sipe04}
\begin{equation}
\Gamma^{ijkl}(\mathbf{r})=\frac{\chi_{3}^{ijkl}(\mathbf{r})}{\epsilon_{0}^{2}n^{8}(\mathbf{r})},
\end{equation}
where $n(\mathbf{r})$ is the position dependent index of refraction. The standard effective nonlinear coefficient $\gamma$ that characterizes the third order nonlinearity is then
\begin{align}
\gamma=&\frac{3\omega_{\text{P}}}{4\varepsilon_{0}^{3}v_{g}^{2}}\nonumber \\
& \times\int\text{d}x\text{d}y\frac{\chi_{3}^{ijmn}\left[d^{i}(x,y)d^{j}(x,y)\right]^{\ast}d^{m}(x,y)d^{n}(x,y)}{n^{8}(x,y)},
\end{align}
where we assume that the transverse mode profiles in the ring are the same as in the channel $\mathcal{D}^i\left(\mathbf{r}\right)=d^i\left(x,y\right)$; here we take $n(x,y)=\bar{n}$ in the waveguide. See Helt et al. \cite{helt12} for details. We take a value of $\gamma=200\text{ (Wm)}^{-1}$, in agreement with recent experiments on ring resonators such as these \cite{azzini12}. For our pump pulses we take bandwidth-limited Gaussians characterized by an intensity FWHM of $\tau_{\text{pump}}$.

\begin{figure}[hbt]
\includegraphics[width=0.45\textwidth]{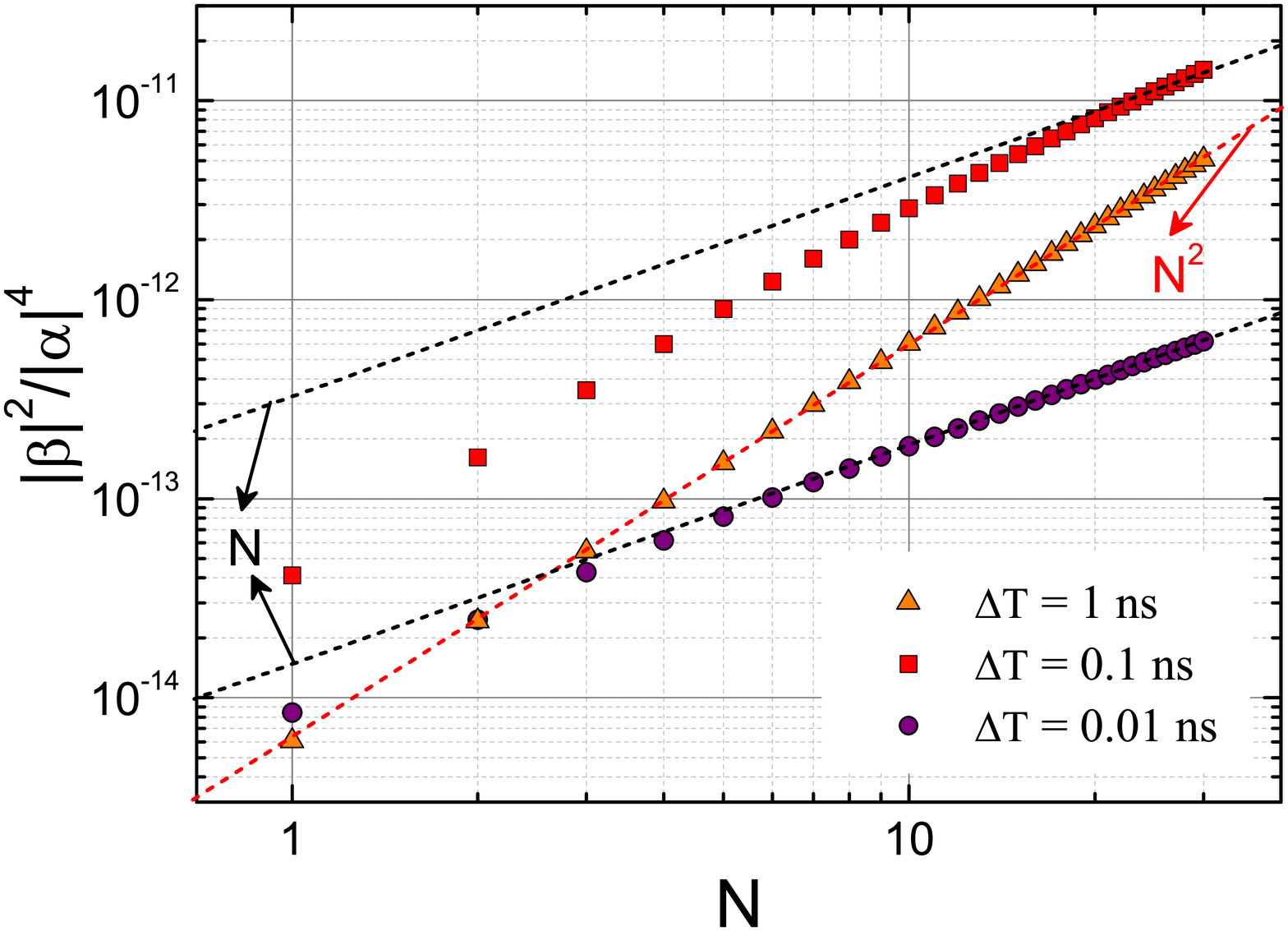}
\caption{Generation efficiency $\left\vert \beta\right\vert ^{2}/\left\vert \alpha\right\vert ^{4}$ vs. ring number.}
\label{fig:ConversionEfficiency}
\end{figure}

We look first at the prediction of the generation efficiency, which can be characterized by the ratio $\left\vert \beta\right\vert ^{2}/\left\vert \alpha\right\vert ^{4}$. We plot this in Fig.~\ref{fig:ConversionEfficiency} as a function of $N$ for different pump pulses. For the longest pump pulse duration considered, $\tau_{\text{pump}}=1\text{ ns}$, we find $N_{\text{coh}}\approx 100$ and we would expect the superradiant $N^{2}$ behavior in generation efficiency to exist for the range of $N$ considered in Fig.~\ref{fig:ConversionEfficiency}; indeed it does. For $\tau_{\text{pump}}=0.1\text{ ns}$ we have $N_{\text{coh}}\approx 10$, leading us to expect that $N\approx 10$ would roughly identify the change of scaling from $N^{2}$ to $N$, and we see from Fig.~\ref{fig:ConversionEfficiency} that it does. For $\tau_{\text{pump}}=0.01\text{ ns}$ the assumption $\delta\ll\Delta$ used in the derivation of \eqref{Ncohresult} formally breaks down, but still the value of $N_{\text{coh}}\approx 1$ would lead us to expect less than superradiant behavior for all $N$, and that is confirmed by the results plotted in Fig.~\ref{fig:ConversionEfficiency}; for $N$ more than a very few the scaling is linear. \textcolor{red}{Finally, we note that while for small $N$ generation efficiencies can be higher for shorter pump pulse durations than for durations in the long pulse limit, they scale differently with increasing N.  Thus, although in our example the generation efficiency in a single ring is smallest for the 1 ns pump pulse, due to the quadratic scaling associated with superradiance this pulse will lead to the largest generation efficiency for larger N.}

\begin{figure}[hbt]
\includegraphics[width=0.45\textwidth]{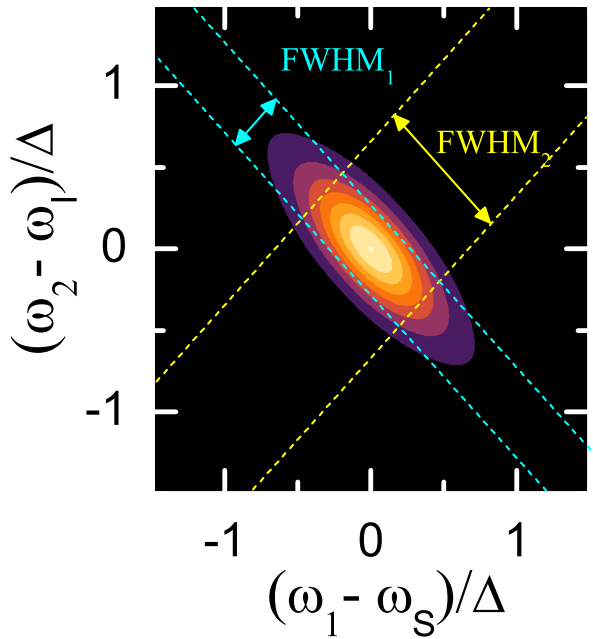}
\caption{The JSD of a single ring for $\tau_{\text{pump}}=0.1\text{ ns}$.}
\label{fig:JSD1}
\end{figure}

While characterizing all of the properties of the generated photon pairs requires the BWF (joint spectral amplitude) $\phi(\omega_{1},\omega_{2})$ \eqref{biphoton}, even the joint spectral density (JSD) $\left\vert \phi(\omega_{1},\omega_{2})\right\vert ^{2}$, where $\left\vert \phi(\omega_{1},\omega_{2})\right\vert ^{2}\text{d}\omega_{1}\text{d}\omega_{2}$ identifies the probability of generating ``photon 1'' within $\text{d}\omega_{1}$ of frequency $\omega_{1}$ and ``photon 2'' within $\text{d}\omega_{2}$ of frequency $\omega_{2}$, contains much information about the quantum correlations of the generated photons. In Fig.~\ref{fig:JSD1} we plot a JSD for a pair generated by SFWM in a single ring resonator with the parameters we have adopted, and subjected to a pump pulse with $\tau_{\text{pump}}=0.1\text{ ns}$, normalizing the frequencies to the line width $\Delta$ of the resonator. The shape of the JSD is essentially determined by the spectral broadness of the pump and the resonance line width. In particular, since the pump pulse duration is longer than the dwelling time $\tau_{\text{dwell}}=$ $0.017\text{ ns}$ in the ring, along the direction $(\omega_{2}-\omega_{\text{I}})=(\omega_{1}-\omega_{\text{S}})$ the FWHM of the JSD (FWHM$_{1}$) is dictated by the frequency width $1/\tau_{\text{pump}}$ of pump pulse through energy conservation, while along $(\omega_{2}-\omega_{\text{I}})=-(\omega_{1}-\omega_{\text{S}})$ the JSD width (FWHM$_{2}$) is determined by the resonance line width $\Delta$, which sets the frequency range in which light can be efficiently generated within the ring \cite{helt11}.

\begin{figure*}[hbt]
\includegraphics[width=0.9\textwidth]{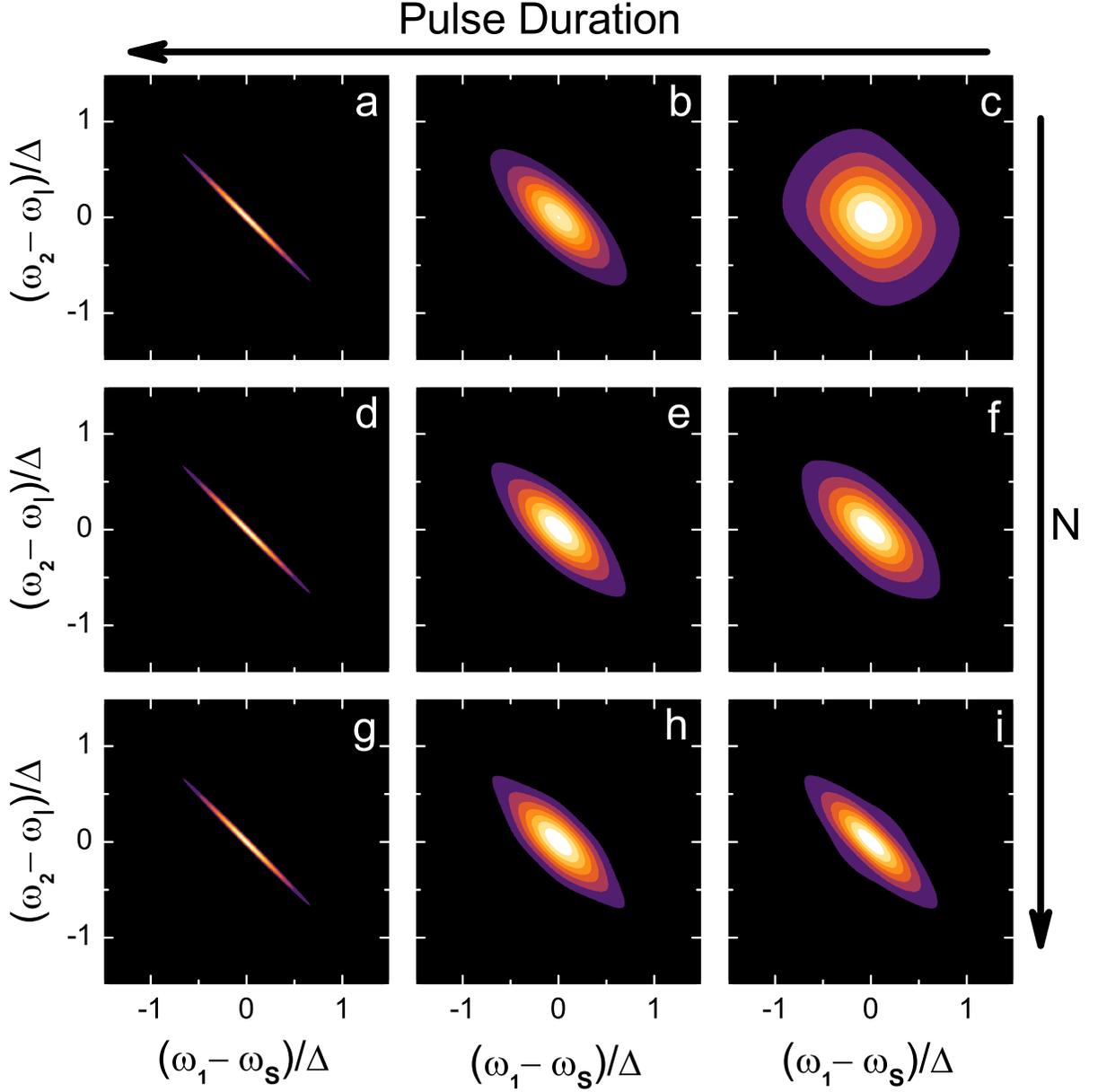}
\caption{JSDs of $N$-ring SCISSORs for various pump pulse durations $\tau_{\text{pump}}$. From top to bottom, $N$ takes the values $1$, $3$, and $5$, while from left to right $\tau_{\text{pump}}$ takes the values 1 ns, 0.1 ns, and 0.01 ns.}
\label{fig:JSDN}
\end{figure*}

It is interesting to consider the JSD for the different scaling regimes identified in Fig.~\ref{fig:ConversionEfficiency}. We consider the examples of $N=1,3,5$. For the pulse durations adopted in Fig.~\ref{fig:ConversionEfficiency}, the results are shown in Fig.~\ref{fig:JSDN}, where we observe that the JSD is in general a function of $N$ and $\tau_{\text{pump}}$. However, as would be expected, in the case of a long pump pulse ($\tau_{\text{pump}}=1\text{ ns}$) (Fig.~\ref{fig:JSDN}a,d, and g) the JSD is essentially independent of $N$; here the structure is fully superradiant for these $N$, and the contribution to the BWF from each ring is essentially identical. There are strong energy correlations due to the narrow spectrum of the pump pulse centered at $\omega_{\text{P}}$, leading to a strong localization of the JSD about the axis $\omega_{1}+\omega_{2}=2\omega_{\text{P}}$, but due to complete constructive interference the contributions to the BWF from each ring are identical and thus no modification of the BWF occurs as more rings are added.

In the case of a short pump pulse ($\tau_{\text{pump}}=0.01\text{ ns}$) (Fig.~\ref{fig:JSDN}c,f,i) we find a significant change in the JSD as $N$ increases. For a single ring (Fig.~\ref{fig:JSDN}c), the JSD has a larger FWHM$_{1}$ than seen for long pump pulses; because of the large spectral width of the pump pulse here, the FWHM$_{1}$ is limited not by that spectral width, but by the resonance line width $\Delta$ of the ring. Yet as we consider larger and larger $N$ (Fig.~\ref{fig:JSDN}f and i), the JSD narrows, indicating stronger energy correlations. This can be understood as due to interference from the contribution of the different rings to the BWF. Along the axis $\omega_{1}+\omega_{2}=2\omega_{\text{P}}$ we have $u=0$ from \eqref{rescond} and \eqref{upsilondef}, and along this line where the energy correlations are strongest we will have $\mu=0$ from \eqref{gammapulse},\eqref{thetapp}, and \eqref{thetasifinal}, and the contributions from the different rings to the full BWF will add in phase. But moving away from this axis we will have $u\neq0$, and thus a nonzero $\mu$ leading to partial destructive interference from the different rings (see \eqref{Jwork}). Thus far from the superradiant regime, as we are here, as $N$ increases there is a significant change in the JSD.

\begin{figure}[hbt]
\includegraphics[width=0.45\textwidth]{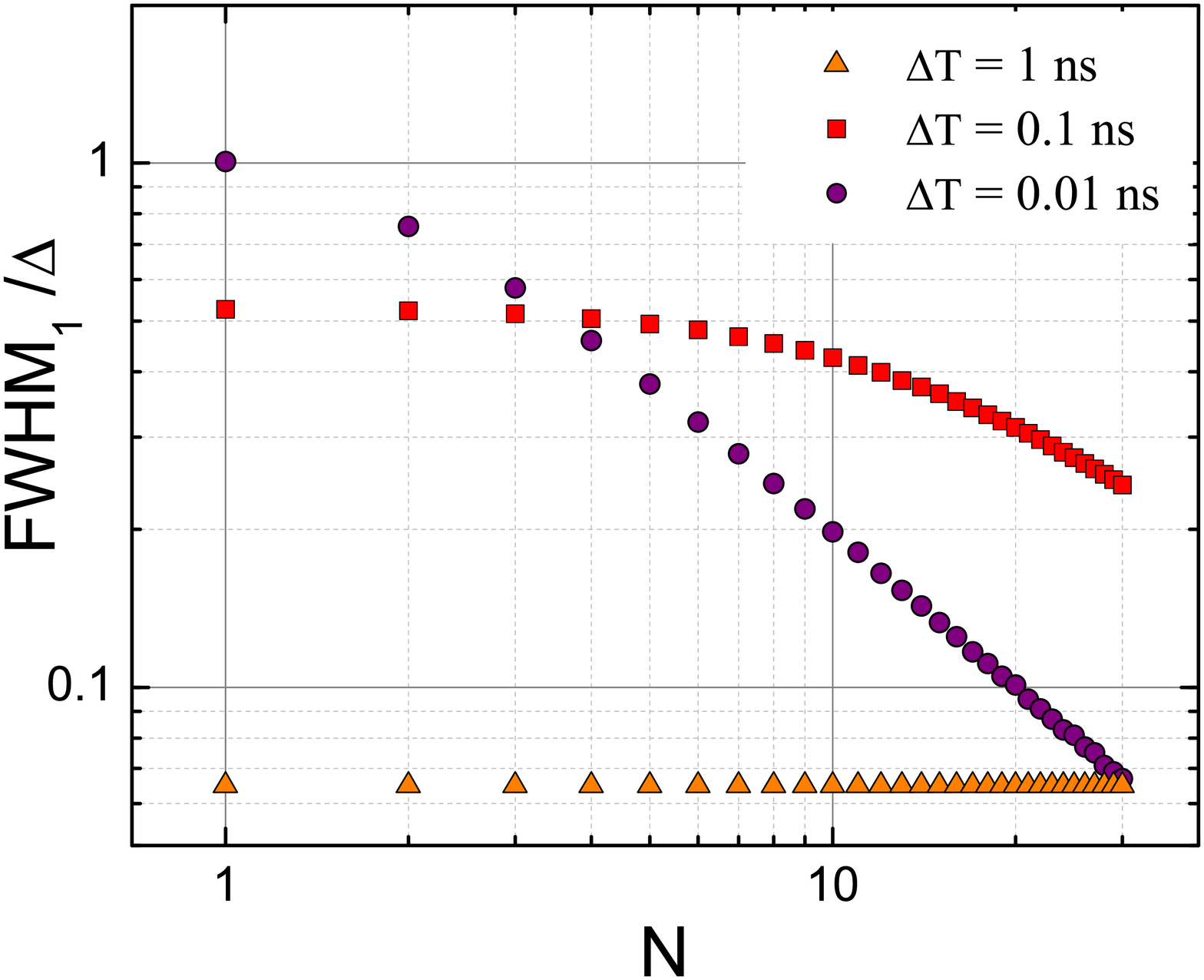}
\caption{FWHM$_{1}$ of the JSD as a function of $N$.}
\label{fig:FWHMN}
\end{figure}

Finally, for an intermediate pump pulse ($\tau_{\text{pump}}=0.1\text{ ns}$) (Fig.~\ref{fig:JSDN}b,e,h) we observe no significant change in the JSD for the small values of $N$ chosen, as the pump pulse duration is sufficiently long that even at $N=5$ we are only just starting to see a decrease in scaling with $N^{2}$ and with it a concomitant slight modification in the BWF, again characterized by stronger energy correlations, as the number of rings increases. The scaling of the FWHM of the JSD along the direction identified by $\omega_{1}-\omega_{\text{S}}=\omega_{2}-\omega_{\text{I}}$ (FWHM$_{1}$) is shown in Fig.~\ref{fig:FWHMN} for a larger range of $N$ than used in Fig.~\ref{fig:JSDN}.

\section{Discussion \label{sec:discussion}}

We have made a number of idealizations in our calculations here. One is that the rings are identical; another is that we have neglected scattering losses in the rings. We plan to turn to including these generalizations, and other extensions of this work, in later publications. We also want to mention that such ``super'' behavior is not really that uncommon in nonlinear optics. Consider for example the second-order classical process of difference frequency generation (DFG) in a channel waveguide. For an injected pump power $P_{\text{P}}$ at $\omega_{\text{P}}$ and an injected signal power $P_{\text{S}}$ at $\omega_{\text{S}}$, an idler power $P_{\text{I}}$ at $\omega_{\text{I}}=\omega_{\text{P}}-\omega_{\text{S}}$ is generated that is given by
\begin{align}
P_{\text{I}}=&P_{\text{S}}\frac{P_{\text{P}}L^{2}}{\mathcal{PA}}\nonumber \\
& \times\text{sinc}^{2}\left\{\left[k_\text{S}(\omega_{\text{S}})+k_\text{I}(\omega_{\text{P}}-\omega_{\text{S}})-k_\text{P}(\omega_{\text{P}})\right]\frac{L}{2}\right\},
\end{align}
where $k_{\text{P},\text{S},\text{I}}(\omega)$ are the wave vector dependences of the pump, signal and idler beams, $\mathcal{A}$ is an effective area of the waveguide, and $\mathcal{P}$ is a quantity with units of power that characterizes the strength of the $\chi_{(2)}$ interaction responsible for the DFG \cite{helt12}. Note that for perfect phase-matching $(k_{\text{S}}(\omega_{\text{S}})+k_{\text{I}}(\omega_{\text{P}}-\omega_{\text{S}})-k_{\text{P}}(\omega_{\text{P}})=0)$ this quantity scales as $L^{2}$, signaling the presence of the coherence of the interaction responsible for the generation over the whole length of the sample, and analogous to the $N^{2}$ dependence in Dicke superradiance or super-SFWM. Even if we consider the spontaneous version of this process, spontaneous parametric down conversion (SPDC), the power generated at idler frequencies is given by
\begin{equation}
P_{\text{I}}=\frac{\hbar(\omega_{\text{P}}/2)}{\mathcal{T}}\frac{P_{\text{P}}L^{2}}{\mathcal{PA}},
\end{equation}
where the leading term can be thought of as the vacuum fluctuations in the signal power leading to the SPDC \cite{helt12}. Here $\mathcal{T}$ is a characteristic time that identifies that power, and depends on the phase-matching condition at different possible idler and signal frequencies; although $\mathcal{T}$ can depend on $L$, that dependence is typically less than linear \cite{helt12}, and so scaling greater than $L$ with channel length can be seen even in this simple structure. \ Similar superlinear scaling is seen in channel waveguides for four-wave-mixing processes, both stimulated and spontaneous.

Nonetheless, we believe the SCISSOR structure we have analyzed here is of particular interest because it points the way to coherently combining contributions to BWFs in integrated structures. One can imagine designing the elements of a structure, and the way in which they are combined, to lead to a greater flexibility than has previously been achieved in determining both the BWFs characterizing the photon pairs generated, as well as the generation efficiency with which they are created. The SCISSOR structure is only a first example, and the simplest of such integrated structures that could be designed and fabricated to lead to new possibilities for integrated sources of quantum correlated light.

\section{Conclusions \label{sec:conclusions}}

\textcolor{red}{In summary, we have investigated how the coherence of multiple quantum nonlinear optical sources pumped by the same field can lead to an alternate form of superradiance than that originally envisioned in a gas of atoms by Dicke~\cite{dicke54}. Focusing on the particular example of a SCISSOR capable of producing pairs of photons via SFWM, we demonstrate a direct analogy with Dicke's work even though the individual microrings of the SCISSOR and spacings between them are both much larger than the wavelengths of light involved.  A sufficiently long exciting pump pulse effectively ``loads'' pump fields in the $N$ rings of the SCISSOR with the correct phase so that the spontaneous generation of signal and idler photons from the different rings adds constructively, leading to super-SFWM -- a quadratic dependence of the generation efficiency on $N$.  We have shown how this coherence follows from a calculation using Fermi's Golden Rule, which is completely analogous to the calculation for Dicke superradiance, and how one can see it in more detail by studying the actual loading of pump fields in the ring by the incident field.  This latter approach allows us to study how the super-SFWM is modified by pump pulses of various duration.}

\textcolor{red}{Additionally, there is a coherence number $N_\text{coh}$ that characterizes the maximum number of rings that can contribute to super-SFWM; beyond this the scaling decreases and approaches a limit that is linearly proportional to $N$.  The coherence number is proportional to the pump pulse length and inversely proportional to the dwelling time of light in a ring.  For $N \lesssim N_\text{coh}$ the biphoton wave function characterizing the spectral features of signal and idler photons generated by the SCISSOR is essentially independent of $N$, since the contributions from each ring add completely coherently.  However, as $N$ increases beyond $N_\text{coh}$ the contributions from all rings add coherently only for very correlated pairs of signal and idler photons, and thus the biphoton wave function exhibits stronger energy correlations.}

\begin{acknowledgments}
This work was supported by the ARC Centre for Ultrahigh bandwidth Devices for Optical Systems (CUDOS)
(Project No. CE110001018), and by the Natural Sciences and Engineering Research Council of Canada (NSERC). L. G. H. acknowledges support from a Macquarie University Research Fellowship.
\end{acknowledgments}

\appendix

\section{Formalism \label{sec:formalism}}

Here we identify the asymptotic-in states for the pump field and the asymptotic-out states for the signal and idler fields. Consider first a single ring (see Fig.~\ref{fig:singleSCISSOR}), which is located just above a bus waveguide, running along the $z$ direction, at $z=0$. An asymptotic-in field corresponding to light incident from the left ($z=-\infty$) with wavenumber $k$ is given by
\begin{equation}
\mathbf{D}_{\text{L}k}^{\text{asy-in}}(\mathbf{r})=\begin{cases}
\frac{\mathbf{d}_{k}(\mathbf{r}_{\perp})e^{ikz}}{\sqrt{2\pi}} & z<0\text{ in the channel},\\
\mathbf{D}_{k}^{\text{ring}}(\mathbf{r}) & \text{inside the ring,}\\
T(k)\frac{\mathbf{d}_{k}(\mathbf{r}_{\perp})e^{ikz}}{\sqrt{2\pi}} & z>0\text{ in the channel},
\end{cases}\label{DLone}
\end{equation}
in the limit of point coupling \cite{heebner04}. Unlike in Section \ref{sec:superradiance}, here we take $k$ to vary continuously over positive values, and in this equation it can range over the frequency range of pump, signal, and idler light. The transverse mode profile in the channel is written as $\mathbf{d}_{k}(\mathbf{r}_{\perp})$ of the field, where $\mathbf{r}_{\perp}=(x,y)$, and we assume that the pump, signal, and idler modes of interest are all associated with the same transverse mode profile; this can easily be generalized. The transmission coefficient $T(k)$ is given \cite{heebner08} by
\begin{equation}
T(k)=\frac{\sigma(k)-e^{ikl}}{1-\sigma(k)e^{ikl}},
\end{equation}
where $l=2\pi R$ is the circumference of the ring, with radius $R$, and we introduce self- and cross-coupling coefficients, $\sigma(k)$ and $\kappa(k)$, in the usual way; we take them to be real, with energy conservation guaranteed by the condition
\begin{equation} 
\sigma^{2}(k)+\kappa^{2}(k)=1.
\end{equation}
The field inside the ring is given by
\begin{equation}
\mathbf{D}_{k}^{\text{ring}}(\mathbf{r})=\frac{i\kappa(k)}{1-\sigma(k)e^{ikl}}\frac{\bm{\mathcal{D}}_{k}(\mathbf{r})}{\sqrt{2\pi}}e^{ik\zeta},\label{Dring}
\end{equation}
where $\zeta$ ranges from $0$ to $l$ and is the distance around the ring starting from $z=0$ (see Fig.~\ref{fig:singleSCISSOR}). In a simple model where the ring and channel have the same dimensions, $\bm{\mathcal{D}}_{k}(\mathbf{r})$ and $\mathbf{d}_{k}(\mathbf{r}_{\perp})$ would describe the same mode profile transverse to the direction of propagation in the ring or channel. Comparing the first and second expressions in \eqref{DLone}, we see that the first factor on the right-hand-side of \eqref{Dring} is a \textit{field enhancement factor,} which identifies how much stronger the field is inside the ring than in the channel. It is useful to express it as a function of frequency, defining
\begin{equation}
\mathfrak{F}(\omega)=\frac{i\kappa\left[k\left(\omega\right)\right]}{1-\sigma\left[k\left(\omega\right)\right]e^{ik(\omega)l}},\label{fieldenhancement}
\end{equation}
where $k(\omega)$ identifies how the wavevector in the channel depends on the frequency. When $k(\omega)l$ is an integer multiple of $2\pi$ the field enhancement factor peaks, as sketched in Fig.~\ref{fig:resonances}; thus this condition identifies the ring resonance, and it is by adjusting the pump frequency to be within one of these resonances, and looking at signal and idler fields at neighboring resonances, that enhancement of spontaneous four-wave mixing occurs \cite{helt12}.

In this simple model the ring is an all-pass structure, and so $\left\vert T(k)\right\vert =1$. With some algebra one can write
\begin{equation}
T(k)=e^{i\theta(k)},
\end{equation}
where 
\begin{equation}
\theta(k)=\pi+kl+2\tan^{-1}\left(\frac{\sigma(k)\sin kl}{1-\sigma(k)\cos kl}\right).\label{thetadef}
\end{equation}
From $\mathbf{D}_{\text{L}k}^{\text{asy-in}}(\mathbf{r})$ it is particularly easy to construct the asymptotic-out field for fields exiting to the right, $\mathbf{D}_{\text{R}k}^{\text{asy-out}}(\mathbf{r})$, since the ring is an all-pass structure ($\left\vert T(k)\right\vert =1$); we have simply
\begin{equation}
\mathbf{D}_{\text{R}k}^{\text{asy-out}}(\mathbf{r})=\frac{1}{T(k)}\mathbf{D}_{\text{L}k}^{\text{asy-in}}(\mathbf{r}),\label{Drelone}
\end{equation}
guaranteeing that we have the desired dependence $\mathbf{d}_{k}(\mathbf{r}_{\perp})e^{ikz}/\sqrt{2\pi}$ for $z>0$ in the channel. We could also construct this by finding first the asymptotic-in field from the right, and using the general relation
\begin{equation}
\left[\mathbf{D}_{\text{R}k}^{\text{asy-out}}(\mathbf{r})\right]^{\ast}=\mathbf{D}_{\text{R}k}^{\text{asy-in}}(\mathbf{r})
\end{equation}
relating asymptotic-in and asymptotic-out fields \cite{liscidini12}.

We now turn to the structure of interest, with $N$ rings spaced such that the minimum distance from the bus waveguide to ring $m$, $m=1,\ldots,N$, occurs at $z_{m}$ (see Fig.~\ref{fig:SCISSOR}). Here the asymptotic-in field from the left is given by
\begin{equation}
\mathbf{D}_{\text{L}k}^{\text{asy-in}}(\mathbf{r})=\begin{cases}
\frac{\mathbf{d}_{k}(\mathbf{r}_{\perp})e^{ikz}}{\sqrt{2\pi}} & z<z_{1}\text{ in the channel},\\
\mathbf{D}_{k}^{\text{ring}}(\mathbf{r};m) & \text{inside ring \ensuremath{m},}\\
\left[T(k)\right]^{m}\frac{\mathbf{d}_{k}(\mathbf{r}_{\perp})e^{ikz}}{\sqrt{2\pi}} & z_{m}<z<z_{m+1}\text{ in the channel},\\
\left[T(k)\right]^{N}\frac{\mathbf{d}_{k}(\mathbf{r}_{\perp})e^{ikz}}{\sqrt{2\pi}} & z>z_{N}\text{ in the channel},
\end{cases}\label{DLmany}
\end{equation}
where
\begin{equation}
\mathbf{D}_{k}^{\text{ring}}(\mathbf{r};m)=e^{ikz_{m}}\left[T(k)\right]^{m-1}\mathbf{D}_{k}^{\text{ring}}(\mathbf{r-}z_{m}\mathbf{\hat{z}}),\label{Dringuse}
\end{equation}
and the function $\mathbf{D}_{k}^{\text{ring}}(\mathbf{r})$ is again given by (\ref{Dring}). In place of \eqref{Drelone}, we have
\begin{equation}
\mathbf{D}_{\text{R}k}^{\text{asy-out}}(\mathbf{r})=\frac{1}{\left[T(k)\right]^{N}}\mathbf{D}_{\text{L}k}^{\text{asy-in}}(\mathbf{r})\label{Drelmany}
\end{equation}
The field operator $\mathbf{D}\left(\mathbf{r}\right)$ can be expanded in terms of either asymptotic-in or asymptotic-out fields; for the quantum states that will be of interest we can restrict ourselves to either using the fields $\mathbf{D}_{\text{L}k}^{\text{asy-in}}(\mathbf{r})$ or the fields $\mathbf{D}_{\text{R}k}^{\text{asy-out}}(\mathbf{r})$; for the frequency range of the pump it is convenient to use the asymptotic-in fields, and for that of the signal and idler the asymptotic-out fields. For the functions $\mathbf{d}_{k}(\mathbf{r})$ properly normalized \cite{yang08} we then have
\begin{align}
\mathbf{D}\left(\mathbf{r}\right)= & \int\text{d}k\,\sqrt{\frac{\hbar\omega(k)}{2}}a_{k}\mathbf{D}_{\text{L}k}^{\text{asy-in}}(\mathbf{r})+\text{H.c.}\nonumber \\
 & +\int\text{d}k\,\sqrt{\frac{\hbar\omega(k)}{2}}b_{k}\mathbf{D}_{\text{R}k}^{\text{asy-out}}(\mathbf{r})+\text{H.c.}\label{Dwrite}
\end{align}
We use $a_{k}$ to indicate lowering operators for modes in the frequency range of the pump resonance, and $b_{k}$ to indicate lower operators for modes in the frequency ranges of both the signal and idler resonances,
\begin{align}
\left[a_{k},a_{k^{\prime}}^{\dagger}\right]= & \left[b_{k},b_{k^{\prime}}^{\dagger}\right]=\delta(k-k^{\prime}),\nonumber \\
\left[a_{k},a_{k^{\prime}}\right]= & \left[b_{k},b_{k^{\prime}}\right]=0,
\end{align}
and all operators associated with pump fields commute with those of the signal and idler fields. We keep implicit the convention that integrals over $k$ are over positive values and those involving pump mode operators range over the frequency range of the pump ring resonance of interest, while those involving signal and idler mode operators range over the frequency ranges of the signal and idler ring resonances of interest.

Returning to the nonlinear Hamiltonian \eqref{Vring}, in terms of both asymptotic-in and -out fields the energy-conserving term involved in spontaneous four-wave mixing is
\begin{align}
V=&-\int\text{d}k_{1}\text{d}k_{2}\text{d}k_{3}\text{d}k_{4}\,S(k_{1},k_{2},k_{3},k_{4})b_{k_{1}}^{\dagger}
b_{k_{2}}^{\dagger}a_{k_{3}}a_{k_{4}}\nonumber \\
& +\text{H.c.},\label{Vdef}
\end{align}
where
\begin{align}
 & S(k_{1},k_{2},k_{3},k_{4})\nonumber \\
 & =\frac{1}{4\epsilon_{0}}\frac{4!}{2!2!}\sqrt{\frac{\hbar\omega(k_{1})\hbar\omega(k_{2})\hbar\omega(k_{3})\hbar\omega(k_{4})}{16}}j(k_{1},k_{2},k_{3},k_{4}),\label{Sdef}
\end{align}
with
\begin{align}
j(k_{1},k_{2},k_{3},k_{4}) =& \int\text{d}\mathbf{r}\;\Gamma_{(3)}^{ijkl}(\mathbf{r})\nonumber \\
& \times\left[D_{\text{R}k_{1}}^{i,\text{asy-out}}(\mathbf{r})D_{\text{R}k_{2}}^{j,\text{asy-out}}(\mathbf{r})\right]^{\ast}\nonumber \\
& \times D_{\text{L}k_{3}}^{k,\text{asy-in}}(\mathbf{r})D_{\text{L}k_{4}}^{l,\text{asy-in}}(\mathbf{r}). \label{jRRLL}
\end{align}
The factor $4!/(2!2!)$ appears in \eqref{Sdef} because we here allow $k_{1}$ and $k_{2}$ to span the ranges associated with both the signal and idler frequency regimes. Now assuming that the nonlinearity is only important in the rings, where the pump field will be strongest because of field enhancement, we can write
\begin{equation}
j(k_{1},k_{2},k_{3},k_{4})=\sum_{m}j^{(m)}(k_{1},k_{2},k_{3},k_{4}),\label{jsum}
\end{equation}
where
\begin{align}
j^{(m)}(k_{1},k_{2},k_{3},k_{4}) =& \left[T(k_{1})T(k_{2})\right]^{N}\int\text{d}\mathbf{r}\,\Gamma_{(3)}^{ijkl}(\mathbf{r})\nonumber \\
& \times\left[D_{k_{1}}^{i,\text{ring}}(\mathbf{r,}m)D_{k_{2}}^{j,\text{ring}}(\mathbf{r},m)\right]^{\ast}\nonumber \\
& \times D_{k_{3}}^{k,\text{ring}}(\mathbf{r},m)D_{k_{4}}^{l,\text{ring}}(\mathbf{r},m),
\end{align}
where we have used \eqref{DLmany} and \eqref{Drelmany}, and the fact that $\left[T^{-1}(k)\right]^{\ast}=T(k)$, which follows since $T(k)$ is of unit norm. The different $j^{(m)}(k_{1},k_{2},k_{3},k_{4})$ differ only by a phase factor, because the rings are physically the same. Using \eqref{Dringuse} we have
\begin{align}
j^{(m)}(k_{1},k_{2},k_{3},k_{4}) =& e^{i(\theta(k_{1})+\theta(k_{2}))N}e^{i(k_{1}+k_{2}-k_{3}-k_{4})z_{1}}\nonumber \\
& \times e^{i\mu(m-1)}j^{\text{ref}}(k_{1},k_{2},k_{3},k_{4}),\label{jmresult}
\end{align}
where recall we have assumed the rings are equally spaced with a distance $\Lambda$ between them,
\begin{equation}
z_{m}-z_{1}=\Lambda(m-1),
\end{equation}
we have put
\begin{align}
\mu=\mu(\left\{ k_{n}\right\} ) = &(k_{3}+k_{4}-k_{1}-k_{2})\Lambda\nonumber \\ 
& +\theta(k_{3})+\theta(k_{4})-\theta(k_{1})-\theta(k_{2}),\label{gamma}
\end{align}
and
\begin{align}
j^{\text{ref}}(k_{1},k_{2},k_{3},k_{4})=&\int\text{d}\mathbf{r}\,\Gamma_{(3)}^{ijkl}(\mathbf{r})\nonumber \\
& \times\left[D_{k_{1}}^{i,\text{ring}}(\mathbf{r})D_{k_{2}}^{j,\text{ring}}(\mathbf{r})\right]^{\ast}\nonumber \\
& \times D_{k_{3}}^{k,\text{ring}}(\mathbf{r})D_{k_{4}}^{l,\text{ring}}(\mathbf{r}),\label{jrefdef}
\end{align}
is independent of $m$; it is to be evaluated for a single ring imagined at $z=0$ above the bus waveguide. The factor $\mu$ contains the effect of the propagation of pump as well as signal and idler light from ring to ring, and the phases acquired by the propagation of pump, signal, and idler through the ring. Using \eqref{jmresult} in \eqref{jsum} and performing the summation, we find
\begin{equation}
j(k_{1},k_{2},k_{3},k_{4})=e^{i\chi}\frac{\sin\frac{\mu N}{2}}{\sin\frac{\mu}{2}}j^{\text{ref}}(k_{1},k_{2},k_{3},k_{4}).\label{jfinal}
\end{equation}
where
\begin{align}
\chi=\chi(\left\{ k_{n}\right\} )=&(k_{3}+k_{4}-k_{1}-k_{2})z_{1}+N\left[\theta(k_{1})+\theta(k_{2})\right]\nonumber \\
& +(N-1)\mu(\left\{ k_{n}\right\} )/2\label{chi}
\end{align}
At this point we can write down the BWF describing the spontaneously created photons. It is usual to move from a $k$ representation to an $\omega$ representation; then the BWF $\phi(\omega_{1},\omega_{2})$, defined for positive $\omega_{1}$ and $\omega_{2}$, is the probability amplitude for ``photon $1$'' being generated within $\text{d}\omega_{1}$ of $\omega_{1}$, and for ``photon $2$'' being generated within $\text{d}\omega_{2}$ of $\omega_{2}$ in the limit of a very low probability that a pair is generated. The BWF is a symmetric function of its arguments, $\phi(\omega_{1},\omega_{2})=\phi(\omega_{2},\omega_{1})$, with both variables $\omega_{1}$ and $\omega_{2}$ ranging over signal and idler frequencies.  It is normalized according to
\begin{equation}
\int\left\vert \phi\left(\omega_{1},\omega_{2}\right)\right\vert ^{2}\text{d}\omega_{1}\text{d}\omega_{2}=1,\label{norm-1}
\end{equation}
and given by \cite{helt12}
\begin{align}
\phi(\omega_{1},\omega_{2})= & \frac{2\sqrt{2}\pi\alpha^{2}}{\beta}\frac{i}{\hbar}\sqrt{\frac{1}{v(\omega_{1})v(\omega_{2})}}\int\text{d}\omega_{3}\int\text{d}\omega_{4}\nonumber \\
& \times\sqrt{\frac{1}{v(\omega_{3})v(\omega_{4})}}\phi_{\text{P}}(\omega_{3})\phi_{\text{P}}(\omega_{4})\nonumber \\
& \times S(k(\omega_{1}),k(\omega_{2}),k(\omega_{3}),k(\omega_{4}))\nonumber \\
& \times \delta(\omega_{1}+\omega_{2}-\omega_{3}-\omega_{4}),\label{BP}
\end{align}
where $v(\omega)=d\omega(k)/dk$ is the group velocity at frequency $\omega$. The pump light is taken to be in a coherent state,
\begin{equation}
\left\langle \mathbf{D}\left(\mathbf{r},t\right)\right\rangle =\alpha\int\text{d}k\,\sqrt{\frac{\hbar\omega(k)}{2}}f_{\text{P}}(k)\mathbf{D}_{\text{L}k}^{\text{asy-in}}(\mathbf{r})e^{-i\omega(k)t}+\text{c.c.} \label{pump}
\end{equation}
in the Heisenberg picture (compare \eqref{Dwrite}); in terms of the complex number $\alpha$, the expectation value of the photon number in the pump pulse is $\left\vert \alpha\right\vert ^{2}$ if the function $f_{\text{P}}(k)$ is normalized,
\begin{equation}
\int\left\vert f_{\text{P}}(k)\right\vert ^{2}\text{d}k=1.\label{fnorm}
\end{equation}
The spectral amplitude $\phi_{\text{P}}(\omega)$,
\begin{equation}
\int\left\vert \phi_{\text{P}}(\omega)\right\vert ^{2}\text{d}\omega=1,\label{pumpnorm}
\end{equation}
is the function of frequency corresponding to $f_{\text{P}}(k)$, where from \eqref{fnorm} and \eqref{pumpnorm} we have $\left\vert f_{\text{P}}(k)\right\vert ^{2}\text{d}k=\left\vert \phi_{\text{P}}(\omega)\right\vert ^{2}\text{d}\omega$. Finally, $\left\vert \beta\right\vert ^{2}$ is the probability per pump pulse that a pair of photons is spontaneously created; it is determined by the expression \eqref{BP} for the BWF and the normalization condition \eqref{norm-1}.

Using \eqref{jfinal} in \eqref{Sdef}, substituting the result in \eqref{BP} and integrating over $\omega_{4}$, we find
\begin{align}
\phi(\omega_{1},\omega_{2})= & \frac{3\pi i\sqrt{2}\alpha^{2}\hbar}{4\varepsilon_{0}\beta}\sqrt{\frac{\omega_{1}\omega_{2}}{v(\omega_{1})v(\omega_{2})}}\int\text{d}\omega\nonumber \\
& \times\sqrt{\frac{\omega(\omega_{1}+\omega_{2}-\omega)}{v(\omega_{p})v(\omega_{1}+\omega_{2}-\omega)}}\phi_{\text{P}}(\omega)\nonumber \\
& \times\phi_{\text{P}}(\omega_{1}+\omega_{2}-\omega)J(\omega_{1},\omega_{2},\omega,\omega_{1}+\omega_{2}-\omega),
\end{align}
where
\begin{align}
J\left(\omega_{1},\omega_{2},\omega_{3},\omega_{4}\right)= & e^{i\chi}\frac{\sin\frac{\mu N}{2}}{\sin\frac{\mu}{2}}\nonumber \\
 & \times j^{\text{ref}}(k(\omega_{1}),k(\omega_{2}),k(\omega_{3}),k(\omega_{4})),\label{JRRLLdef}
\end{align}
we use $k(\omega)$ again to indicate the wave number written as a function of frequency, and we now consider $\mu$ and $\chi$ to depend on the frequencies $\left\{ \omega_{n}\right\} $ through the dependence of the $\left\{ k_{n}\right\} $ on frequency; from (\ref{gamma}), (\ref{chi}) we now take
\begin{align}
\mu= & \mu\left(\left\{ \omega_{n}\right\} \right)\nonumber \\
\equiv & \left[k(\omega_{3})+k(\omega_{4})-k(\omega_{1})-k(\omega_{2})\right]\Lambda\nonumber \\
 & +\theta\left[k(\omega_{3})\right]+\theta\left[k(\omega_{4})\right]-\theta\left[k(\omega_{1})\right]-\theta\left[k(\omega_{2})\right],\label{gammageneral}
\end{align}
\begin{align}
\chi= & \chi\left(\left\{ \omega_{n}\right\} \right)\nonumber \\
\equiv & \left[k(\omega_{3})+k(\omega_{4})-k(\omega_{1})-k(\omega_{2})\right]z_{1}\nonumber \\
 & +N\left\{ \theta\left[k(\omega_{1})\right]+\theta\left[k(\omega_{2})\right]\right\} +(N-1)\mu/2.\label{chigeneral}
\end{align}

\section{Long pulse limit details \label{sec:longpulsedetails}}

In this Appendix we construct the limit of the expression \eqref{betasquaredgeneral} for $\left\vert \beta\right\vert ^{2}$ assuming a very long pump pulse. Taking \eqref{biphoton} we note that for $\phi_{\text{P}}(\omega)$ strongly peaked at $\omega=\omega_{\text{P}}$ and $\phi_{\text{P}}(\omega_{1}+\omega_{2}-\omega)$ strongly peaked at $\omega_{1}+\omega_{2}-\omega=\omega_{\text{P}}$ by the form \eqref{phipulse} adopted, we have
\begin{align}
\phi(\omega_{1},\omega_{2})\approx & \frac{3\pi i\sqrt{2}\alpha^{2}\hbar}{4\varepsilon_{0}\beta}\sqrt{\frac{\omega_{1}\omega_{2}}{v(\omega_{1})v(\omega_{2})}}\frac{\omega_{\text{P}}}{v(\omega_{\text{P}})}\nonumber \\
& \times J(\omega_{1},\omega_{2},\omega_{\text{P}},\omega_{\text{P}})\int\text{d}\omega\,\phi_{\text{P}} (\omega) \nonumber \\
& \times\phi_{\text{P}}(\omega_{1}+\omega_{2}-\omega).
\end{align}
Using the spectral amplitude \eqref{phipulse} the integral over $\omega$ can be performed using the identity
\begin{equation}
\text{sinc}(y-z)=\frac{1}{\pi}\int\text{d}x\,\text{sinc}(x-y)\text{ sinc}(x-z),
\end{equation}
and we find
\begin{align}
\phi\left(\omega_{1},\omega_{2}\right)\approx & \frac{3\pi i\sqrt{2}\alpha^{2}\hbar}{4\varepsilon_{0}\beta}\sqrt{\frac{\omega_{1}\omega_{2}}{v(\omega_{1})v(\omega_{2})}}\frac{\omega_{\text{P}}}{v(\omega_{\text{P}})}\nonumber \\
 & \times J(\omega_{1},\omega_{2},\omega_{\text{P}},\omega_{\text{P}})\text{sinc}\left(\frac{2\omega_{\text{P}}-\omega_{1}-\omega_{2}}{\Delta\omega}\right).\label{philongwork}
\end{align}
Within these approximations we determine $\left\vert \beta\right\vert ^{2}$ from the normalization condition \eqref{norm}. \ For $\Delta\omega$ small enough, in the expression that results we can put
\begin{equation}
\frac{1}{\pi\Delta\omega}\text{sinc}^{2}\left(\frac{2\omega_{\text{P}}-\omega_{1}-\omega_{2}}{\Delta\omega}\right)\rightarrow\delta(2\omega_{\text{P}}-\omega_{1}-\omega_{2}),
\end{equation}
and thus the double integral in \eqref{norm} reduces to a single integral, giving
\begin{align}
\left\vert \beta\right\vert ^{2}\approx &\frac{2\pi}{\Delta T}\left\vert \frac{3\pi\sqrt{2}\alpha^{2}\hbar}{4\varepsilon_{0}}\right\vert ^{2}\frac{\omega_{\text{P}}^{2}}{v^{2}(\omega_{\text{P}})}\int \text{d}\omega_{1}\frac{\omega_{1}(2\omega_{\text{P}}-\omega_{1})}{v(\omega_{1})v(2\omega_\text{{P}}-\omega_{1})}\nonumber \\
& \times\left\vert J(\omega_{1},2\omega_{\text{P}}-\omega_{1},\omega_{\text{P}},\omega_{\text{P}})\right\vert ^{2}.
\end{align}
Recall that $\omega_{1}$ ranges over frequencies of the signal \textit{and} idler. Because of the symmetry of the integrand, we can replace this by an integral that ranges over only the frequency range of the signal, if we simply multiply by $2$. This yields \eqref{betasquared1}. Within the same approximation the joint spectral density $\left\vert \phi(\omega_{1},\omega_{2})\right\vert ^{2}$ that follows from (\ref{philongwork}) is immediately seen to be \eqref{JSDlong}.

\section{Connection to Fermi's Golden Rule \label{sec:connection}}

To connect the result for the photon generation rate in the long pulse limit \eqref{betasquared2} to the simple result of Fermi's Golden Rule \eqref{ringresult} we need to relate the $\alpha_{m}$ \eqref{alpham} associated with the individual rings to the amplitude $\alpha$ characterizing the photon number in the pump pulse appearing in (\ref{pump}). That is, we must identify how the incident pump pulse effectively ``loads'' each ring with pump light. \ And to do that we must first relate the mode field $\mathbf{D}_{\text{iso}}(\mathbf{r},m)$ of \eqref{DP} characterizing an isolated ring $m$ to the mode field $\mathbf{D}_{k}^{\text{ring}}(\mathbf{r};m)$ characterizing ring $m$ in the asymptotic-in state (\ref{DLmany}). For the convention adopted in \eqref{DP}, $\mathbf{D}_{\text{iso}}(\mathbf{r},m)$ satisfies \cite{yang08} the normalization condition
\begin{equation}
\int_{V_{m}}\frac{\mathbf{D}_{\text{iso}}^{\ast}(\mathbf{r},m)\cdot\mathbf{D}_{\text{iso}}(\mathbf{r},m)}{\varepsilon_{0}\varepsilon(\mathbf{r})}\text{d}\mathbf{r}=1,\label{isonorm}
\end{equation}
where the integral is over the $m^{\text{th}}$ ring. In the approximation \eqref{DP} of isolated rings the pump frequency in the ring is $\omega_{\text{P}}$, and letting $k_{\text{P}}$ identify $k(\omega_{\text{P}})$ we can write the field associated with ring $m$ in the asymptotic-in state at $\omega_{\text{P}}$ (see \eqref{Dringuse}) as
\begin{align}
\mathbf{D}_{k_{\text{P}}}^{\text{ring}}(\mathbf{r};m)= & e^{ik_{\text{P}}z_{m}}\left[T(k_{\text{P}})\right]^{m-1}\mathbf{D}_{k_{\text{P}}}^{\text{ring}}(\mathbf{r-}z_{m}\mathbf{\hat{z}})\nonumber \\
= & e^{ik_{\text{P}}z_{m}}\left[T(k_{\text{P}})\right]^{m-1}\mathfrak{F}(\omega_{\text{P}})\nonumber \\
& \times\frac{\bm{\mathcal{D}}_{k_{\text{P}}}(\mathbf{r}-z_{m}\mathbf{\hat{z}})}{\sqrt{2\pi}}e^{ik_{\text{P}}\zeta},
\end{align}
where we have used \eqref{Dring} and the field enhancement factor $\mathfrak{F}(\omega_{\text{P}})$ is given by \eqref{fieldenhancement}.

We want to identify the proportionality between $\mathbf{D}_{k_{\text{P}}}^{\text{ring}}(\mathbf{r};m)$ and $\mathbf{D}_{\text{iso}}(\mathbf{r},m)$, putting $\mathbf{D}_{k_{\text{P}}}^{\text{ring}}(\mathbf{r};m)=C\mathbf{D}_{\text{iso}}(\mathbf{r},m)$, where $C$ is a constant to be determined. We integrate $\left[\mathbf{D}_{k_{\text{P}}}^{\text{ring}}(\mathbf{r};m)\right]^{\ast}\cdot\mathbf{D}_{k_{\text{P}}}^{\text{ring}}(\mathbf{r};m)/\left(\varepsilon_{0}\varepsilon(\mathbf{r})\right)$ over the $m^{\text{th}}$ ring; that integral will involve an integral over the length of the ring, which will give a factor $l=2\pi R$ with $l$ the circumference of the ring and $R$ its radius, as well as an integral over the plane perpendicular to the direction of propagation of the light in the ring. \ Recalling that we have taken $\mathcal{D}_{k_{\text{P}}}(\mathbf{r})$ to describe the same physical mode profile as $\mathbf{d}_{k_{\text{P}}}(\mathbf{r}_{\bot})$ (see text after \eqref{Dring}), we can use the normalization condition \cite{yang08} of $\mathbf{d}_{k}(\mathbf{r}_{\perp})$,
\begin{equation}
\int\frac{\mathbf{d}_{k}^{\ast}(\mathbf{r}_{\perp})\cdot\mathbf{d}_{k}(\mathbf{r}_{\perp})}{\epsilon_{0}\varepsilon(\mathbf{r})}\text{d}x\text{d}y=1,
\end{equation}
to complete the integral of $\left[\mathbf{D}_{k_{\text{P}}}^{\text{ring}}(\mathbf{r};m)\right]^{\ast}\cdot\mathbf{D}_{k_{\text{P}}}^{\text{ring}}(\mathbf{r};m)/\left(\varepsilon_{0}\varepsilon(\mathbf{r})\right)$ over the ring. Using \eqref{isonorm} we can then identify the constant $C$, within a phase, and write
\begin{equation}
\mathbf{D}_{k_{\text{P}}}^{\text{ring}}(\mathbf{r};m)=e^{ik_{\text{P}}z_{m}}\left[T(k_{\text{P}})\right]^{m-1}\mathfrak{F}(\omega_{\text{P}})\sqrt{R}\;\mathbf{D}_{\text{iso}}(\mathbf{r},m).\label{asymiso}
\end{equation}
The phase chosen is (except for perhaps a global phase factor) the natural one, since it is natural to take $\mathbf{D}_{\text{iso}}(\mathbf{r},m)=\mathbf{D}_{\text{iso}}(\mathbf{r}-z_{m}\mathbf{\hat{z}})$, where $\mathbf{D}_{\text{iso}}(\mathbf{r})$ would be the mode field of an isolated ring with its closest point to the bus waveguide at $z=0$; that is, the $\mathbf{D}_{\text{iso}}(\mathbf{r},m)$ are essentially all the same, except for the position of the ring, and so the phase factors associated with the propagation of the asymptotic-in state will reside in $\mathbf{D}_{k_{\text{P}}}^{\text{ring}}(\mathbf{r};m)$.

We can now identify the $\alpha_{m}$ of \eqref{alpham} that would be associated with the pump field in each ring for excitation by a long pulse characterized by $\alpha$. From \eqref{pump} we have
\begin{align}
\left\langle \mathbf{D}\left(\mathbf{r},t\right)\right\rangle =&\alpha\int\frac{\text{d}k}{\text{d}\omega}\sqrt{\frac{\hbar\omega}{2}}\sqrt{\frac{\text{d}\omega}{\text{d}k}}\phi_{\text{P}}(\omega)\nonumber \\
& \times\mathbf{D}_{\text{L}k(\omega)}^{\text{asy-in}}(\mathbf{r})e^{-i\omega t}\text{d}\omega+\text{c.c.},
\end{align}
where we have converted to an integral over frequency and used $\left\vert f_{\text{P}}(k)\right\vert ^{2}\text{d}k=\left\vert \phi_{\text{P}}(\omega)\right\vert ^{2}\text{d}\omega$ (see text after \eqref{pumpnorm}). For a $\phi_{\text{P}}(\omega)$ strongly peaked at $\omega_{\text{P}}$ we can write $\mathbf{D}_{\text{L}k(\omega)}^{\text{asy-in}}(\mathbf{r})\approx\mathbf{D}_{\text{L}k_{\text{P}}}^{\text{asy-in}}(\mathbf{r})$, and put $\text{d}\omega/\text{d}k\approx v(\omega_{\text{P}})$ and $\omega\approx\omega_{\text{P}}$ in the prefactors of $\phi_{\text{P}}(\omega)$ to write
\begin{align}
\left\langle \mathbf{D}\left(\mathbf{r},t\right)\right\rangle \approx & \alpha\sqrt{\frac{\pi\hbar\omega_{\text{P}}}{v(\omega_{\text{P}})}}\mathbf{D}_{\text{L}k_{\text{P}}}^{\text{asy-in}}(\mathbf{r})\int\phi_{\text{P}}(\omega)e^{-i\omega t}\frac{\text{d}\omega}{\sqrt{2\pi}}+\text{c.c.}\nonumber \\
= & \alpha\sqrt{\frac{\pi\hbar\omega_{\text{P}}}{v(\omega_{\text{P}})}}\mathbf{D}_{\text{L}k_{\text{P}}}^{\text{asy-in}}(\mathbf{r})\frac{e^{-i\omega_{\text{P}}t}}{\sqrt{\Delta T}}+\text{c.c.}\nonumber \\
& \quad\text{for }-\frac{\Delta T}{2}<t<\frac{\Delta T}{2},\nonumber \\
= & 0\text{ otherwise,}
\end{align}
where in the last two lines we have specialized to the pump pulse of \eqref{tophat}. Now using the expression \eqref{DLmany} for $\mathbf{D}_{\text{L}k}^{\text{asy-in}}(\mathbf{r})$, and \eqref{asymiso} for the field in the ring, we find 
\begin{align}
\left\langle \mathbf{D}\left(\mathbf{r},t\right)\right\rangle _{\text{ring}\text{ }m}\approx & \alpha\sqrt{\frac{\pi R\hbar\omega_{\text{P}}}{v(\omega_{\text{P}})\left(\Delta T\right)}}e^{ik_{\text{P}}z_{m}}\left[T(k_{\text{P}})\right]^{m-1}\mathfrak{F}(\omega_{\text{P}})\nonumber \\
& \times\mathbf{D}_{\text{iso}}(\mathbf{r},m)e^{-i\omega_{\text{P}}t}+\text{c.c.}\nonumber \\
& \quad\text{for }-\frac{\Delta T}{2}<t<\frac{\Delta T}{2}\nonumber \\
\approx & 0\text{ otherwise.}
\end{align}
Since from the expression \eqref{DP} for a pump field loaded into the rings we would have
\begin{equation}
\left\langle \mathbf{D}\left(\mathbf{r},t\right)\right\rangle _{\text{ring}\text{ }m}=\sum_{m}\sqrt{\frac{\hbar\omega_{\text{P}}}{2}}\alpha_{m}\mathbf{D}_{\text{iso}}(\mathbf{r},m)e^{-i\omega_{\text{P}}t}+\text{c.c.},
\end{equation}
for the time $\Delta T$ that we were making a calculation using Fermi's Golden Rule, we can identify the relation
\begin{equation}
\alpha_{m}=\alpha\sqrt{\frac{l}{v(\omega_{\text{P}})\left(\Delta T\right)}}e^{ik_{\text{P}}z_{m}}\left[T(k_{\text{P}})\right]^{m-1}\mathfrak{F}(\omega_{\text{P}}),\label{alphamalpha}
\end{equation}
where again $l=2\pi R$ is the ring length and $R$ the ring radius.

To now make the connection of the result \eqref{betasquared2} with the calculation using Fermi's Golden Rule, we look at
\begin{align}
\alpha^{2}J(\omega_{1},\omega_{2},\omega_{\text{P}},\omega_{\text{P}})= & \alpha^{2}\sum_{m}\int_{V_{m}}\text{d}\mathbf{r}\,\Gamma_{(3)}^{ijkl}(\mathbf{r})\nonumber \\
& \times\left[D_{\text{R}k_{1}}^{i,\text{asy-out}}(\mathbf{r})D_{\text{R}k_{2}}^{j,\text{asy-out}}(\mathbf{r})\right]^{\ast}\nonumber \\
& \times D_{\text{L}k_{\text{P}}}^{k,\text{asy-in}}(\mathbf{r})D_{\text{L}k_{\text{P}}}^{l,\text{asy-in}}(\mathbf{r}),
\end{align}
where we have used \eqref{JRRLLdef} together with \eqref{jfinal} and the definition \eqref{jRRLL}; here $k_{1}=k(\omega_{1})$ and $k_{2}=k(\omega_{2})$, the sum is over all the rings, and the integration over volume $V_{m}$ over is over ring $m$. Within ring $m$ we have $\mathbf{D}_{\text{L}k_{\text{P}}}^{\text{asy-in}}(\mathbf{r})$ equal to $\mathbf{D}_{k_{\text{P}}}^{\text{ring}}(\mathbf{r};m)$, and using the result \eqref{asymiso} we have
\begin{align}
&\alpha^{2}J(\omega_{1},\omega_{2},\omega_{\text{P}},\omega_{\text{P}})\nonumber \\
&= \alpha^{2}\sum_{m}e^{2ik_{\text{P}}z_{m}}\left[T(k_{\text{P}})\right]^{2(m-1)}\mathfrak{F}^{2}(\omega_{\text{P}})R\int_{V_{m}}\text{d}\mathbf{r}\,\Gamma_{(3)}^{ijkl}(\mathbf{r})\nonumber \\
& \quad\times\left[D_{\text{R}k_{1}}^{i,\text{asy-out}}(\mathbf{r})D_{\text{R}k_{2}}^{j,\text{asy-out}}(\mathbf{r})\right]^{\ast}D_{\text{iso}}^{k}(\mathbf{r,}m)D_{\text{iso}}^{l}(\mathbf{r},m)\nonumber \\
&= \alpha^{2}\sum_{m}e^{2ik_{\text{P}}z_{m}}\left[T(k_{\text{P}})\right]^{2(m-1)}\mathfrak{F}^{2}(\omega_{\text{P}})R F^{(m)}(\omega_{1},\omega_{2})\nonumber \\
&= \frac{v(\omega_{\text{P}})(\Delta T)}{2\pi}\sum_{m}\alpha_{m}^{2}F^{(m)}(\omega_{1},\omega_{2}),\label{asJ}
\end{align}
where in the third line we have recalled the definition \eqref{Fdef} of $F^{(m)}(\omega_{1},\omega_{2})$, and in the fourth line we have used the relation \eqref{alphamalpha} between $\alpha_{m}$ and $\alpha$. \ Using (\ref{asJ}) in \eqref{betasquared2} we find
\begin{align}
\frac{\left\vert \beta\right\vert ^{2}}{\Delta T}\approx& \frac{9\pi\hbar^{2}\omega_{\text{P}}^{2}}{8\varepsilon_{0}^{2}}\int\text{d}\omega\frac{\omega(2\omega_{\text{P}}-\omega)}{v(\omega)v(2\omega_{\text{P}}-\omega)}\nonumber \\
& \times\left\vert \sum_{m}\alpha_{m}^{2}F^{(m)}(\omega,2\omega_{P}-\omega)\right\vert ^{2}.\label{limit}
\end{align}
To compare this with the result \eqref{ringresult} from Fermi's Golden Rule, in that latter result we pass to an infinite normalization length $\mathcal{L}$; the distance between the rings and the number of rings of course remains fixed; then $k_{s}$ and $k_{i}$ become continuous variables, $\sum_{k_{s},k_{i}}\rightarrow\left(\mathcal{L}/2\pi\right)^{2}\int \text{d}k_{1}\text{d}k_{2}$, and integrating over all possible final states we have
\begin{align}
\frac{\mathfrak{P}(t)}{t}=&\frac{9\pi\hbar^{3}\omega_{\text{P}}^{2}}{8\varepsilon_{0}^{2}}\int\text{d}k_{1}\text{d}k_{2}\,\omega_{1}\omega_{2}\left\vert \sum_{m}\alpha_{m}^{2}F^{(m)}(\omega_{1},\omega_{2})\right\vert^{2}\nonumber \\
& \times\delta(\hbar\omega_{1}+\hbar\omega_{2}-2\hbar\omega_{\text{P}}).
\end{align}
Changing the integrals to those over frequency, and using $\text{d}k/\text{d}\omega=1/v(\omega)$, where $v(\omega)$ is the group velocity, we can integrate over $\text{d}\omega_{2}$ with the use of the Dirac delta function and find
\begin{align}
\frac{\mathfrak{P}(t)}{t}=&\frac{9\pi\hbar^{2}\omega_{\text{P}}^{2}}{8\varepsilon_{0}^{2}}\int\text{d}\omega\frac{\omega(2\omega_{\text{P}}-\omega)}{v(\omega)v(2\omega_{\text{P}}-\omega)}\nonumber \\
& \times\left\vert \sum_{m}\alpha_{m}^{2}F^{(m)}(\omega,2\omega_{\text{P}}-\omega)\right\vert ^{2},
\end{align}
in agreement with \eqref{limit}.

\section{The coherence number \label{sec:coherencenumber}}

With frequency ranges restricted according to~\eqref{pumpconstraints} and $\left\vert u\right\vert\lesssim\delta$ (recall~\eqref{upsilondef}), here we calculate a useful approximation for the number of rings that will behave coherently.  We first rewrite the single ring component~\eqref{jrefwork} of $J(\omega_{1},\omega_{2},\omega,\omega_{1}+\omega_{2}-\omega)$ as
\begin{align}
 & j^{\text{ref}}\left[k(\omega_{1}),k(\omega_{2}),k(\omega),k(\omega_{1}+\omega_{2}-\omega)\right]\nonumber \\
 & =\left(\frac{2}{1-\sigma}\right)^{2}\frac{\Delta/2}{(\omega_{1}-\omega_{\text{S}})-i(\Delta/2)}\frac{\Delta/2}{(\omega_{2}-\omega_{\text{I}})-i(\Delta/2)}\nonumber \\
 & \quad\times\frac{\Delta/2}{(\omega-\omega_{\text{P}})-i(\Delta/2)}\frac{\Delta/2}{(\omega-\omega_{1}-\omega_{2}-\omega_{\text{P}})-i(\Delta/2)}\nonumber \\
 & \quad\times\int\frac{\text{d}\mathbf{r}}{(2\pi)^{2}}\Gamma_{(3)}^{ijkl}(\mathbf{r})\left[\mathcal{D}^{i}(\mathbf{r})\mathcal{D}^{j}(\mathbf{r})\right]^{\ast}\mathcal{D}^{k}(\mathbf{r})\mathcal{D}^{l}(\mathbf{r}), \label{jrefresult}
\end{align}
showing the expected resonance behavior. From the general expression \eqref{JRRLLdefmaintext} for $J(\omega_{1},\omega_{2},\omega_{3},\omega_{4})$ we now seek
\begin{align}
 & J(\omega_{1},\omega_{2},\omega,\omega_{1}+\omega_{2}-\omega)\nonumber \\
 & =e^{i\chi}\frac{\sin\frac{\mu N}{2}}{\sin\frac{\mu}{2}}j^{\text{ref}}(k(\omega_{1}),k(\omega_{2}),k(\omega),k(\omega_{1}+\omega_{2}-\omega)), \label{Jwork}
\end{align}
where the general expression \eqref{gammageneral} for $\mu$ reduces, in the limit \eqref{noGVD} of vanishing group velocity dispersion to (recall \eqref{rescond} and \eqref{energyconservation})
\begin{equation}
\mu=\theta\left[k(\omega)\right]+\theta\left[k(\omega_{1}+\omega_{2}-\omega)\right]-\theta\left[k(\omega_{1})\right]-\theta\left[k(\omega_{2})\right],\label{gammapulse}
\end{equation}
and the general expression \eqref{chigeneral} for $\chi$ reduces in that same limit to
\begin{equation}
\chi=N\left\{ \theta\left[k(\omega_{1})\right]+\theta\left[k(\omega_{2})\right]\right\} +\mu(N-1)/2.\label{chipulse}
\end{equation}
Here, for simplicity, we have put $z_{1}=0$ and, in both of these expressions (see \eqref{thetadef}),
\begin{equation}
\theta(k)=\pi+kl+2\tan^{-1}\left(\frac{\sigma\sin kl}{1-\sigma\cos kl}\right).
\end{equation}
In the limit of weak coupling \eqref{WC} and for small deviations in $k\left(\omega\right)$ from a resonance, we simplify this as
\begin{equation}
\theta(k)\approx\pi+kl+2\tan^{-1}\left(\frac{\sin kl}{1-\sigma}\right).
\end{equation}
Then, again neglecting group velocity dispersion and Taylor expanding $k\left(\omega\right)$ to first order \eqref{noGVD}, as well as recalling the resonance condition \eqref{rfrequencies}, we write
\begin{align}
\theta\left[k(\omega)\right]\approx &\left(2M+1\right)\pi+\frac{(\omega-\omega_{\text{M}})l}{v_{g}}\nonumber \\
&+2\tan^{-1}\left(\frac{2(\omega-\omega_{\text{M}})}{\Delta}\right).
\end{align}
For the first two terms of \eqref{gammapulse}, the pump terms, $\delta\ll\Delta$ guarantees that the argument of $\tan^{-1}$ is small and we have
\begin{align}
& \theta\left[k\left(\omega\right)\right]+\theta\left[k\left(\omega_{1}+\omega_{2}-\omega\right)\right]\nonumber \\
& \approx2\pi\left(2P+1\right)+\frac{ul}{v_{g}}+\frac{4u}{\Delta}. \label{thetapp}
\end{align}
For the second two terms of \eqref{gammapulse}, the generated terms, we cannot simplify the $\tan^{-1}$ functions the same way, since from the resonant behavior in \eqref{jrefresult} we must consider $\omega_{1}-\omega_{\text{S}}$ and $\omega_{2}-\omega_{\text{I}}$ ranging over $\Delta$. However, using the identity $\tan^{-1}\left(x\right)+\tan^{-1}\left(y\right)=\tan^{-1}\left[\left(x+y\right)/\left(1-xy\right)\right]$ we find
\begin{align}
 & \theta\left[k\left(\omega_{1}\right)\right]+\theta\left[k\left(\omega_{2}\right)\right]\nonumber  \\
 & \approx2\pi(S+I+1)+\frac{ul}{v_{g}}+2\tan^{-1}\left(\frac{2u}{\Delta+\frac{\eta^{2}-u^{2}}{\Delta}}\right), \label{thetasi}
\end{align}
where we have defined
\begin{equation}
\eta=(\omega_{1}-\omega_{\text{S}})-(\omega_{2}-\omega_{\text{I}}).
\end{equation}
The denominator in the $\tan^{-1}$ function of (\ref{thetasi}) is always positive and of order $\Delta$ or larger, since $\left\vert u\right\vert $ is of order $\delta$ and $\delta\ll\Delta$, and so the argument of that function is always much less than unity; keeping the first term in the expansion we write
\begin{align}
\theta\left[k\left(\omega_{1}\right)\right]+\theta\left[k\left(\omega_{2}\right)\right]\approx & 2\pi(S+I+1)+\frac{ul}{v_{g}}\nonumber \\
&+\frac{4u}{\Delta}\frac{\Delta^{2}}{\Delta^{2}+\eta^{2}-u^{2}}.\label{thetasifinal}
\end{align}
For a pump pulse with $\left\vert \phi_{\text{P}}(\omega)\right\vert ^{2}$ peaked at the ring resonance frequency $\omega_{\text{P}}$, energy conservation and the resonance structure of $j^{\text{ref}}$ \eqref{jrefresult} will allow for a range of $\omega_{1}$ and $\omega_{2}$ with $\eta$ varying over a range of about $-\Delta/2$ to $\Delta/2$. For most of this range we will have $\eta^{2}\gg u^{2}$, so as a rough estimate we neglect $u^{2}$ in the denominator of the last term in \eqref{thetasifinal} and approximate $\Delta^{2}/(\Delta^{2}+\eta^{2})\approx1/2$. Then taking a typical value of $\left\vert u\right\vert $ to be about $\delta$, we characterize the typical size of the expressions in \eqref{thetapp} and \eqref{thetasifinal} to be
\begin{align}
& \theta\left[k\left(\omega\right)\right]+\theta\left[k\left(\omega_{1}+\omega_{2}-\omega\right)\right]\nonumber \\
& \approx2\pi\left(2P+1\right)+\left(\frac{l}{v_{g}}+\frac{4}{\Delta}\right)\delta,
\end{align}
and
\begin{align}
&\theta\left[k\left(\omega_{1}\right)\right]+\theta\left[k\left(\omega_{2}\right)\right]\nonumber \\
&\approx2\pi(S+I+1)+\left(\frac{l}{v_{g}}+\frac{2}{\Delta}\right)\delta\label{approxthetas}
\end{align}
respectively.  From \eqref{gammapulse}, this gives a typical size for $\mu$ of about~\eqref{eq:mumaintext}. We also note that a typical size of the phase $\chi$ appearing in \eqref{Jwork} for $J(\omega_{1},\omega_{2},\omega,\omega_{1}+\omega_{2}-\omega)$, given by \eqref{chipulse}, is found from \eqref{approxthetas} and \eqref{eq:mumaintext} as
\begin{align}
\chi\approx & 2\pi N(S+I+1)+\left(N\left(\frac{l}{v_{g}}+\frac{2}{\Delta}\right)+\frac{N-1}{\Delta}\right)\delta\nonumber \\
\approx & 2\pi N(S+I+1)+\left(\frac{3N\delta}{\Delta}\right),
\end{align}
where in the second line we have used our assumption that $\delta/\Delta\ll1$ and have taken the finesse of the rings to be very high, so that the free spectral range (\ref{FSR}) is much larger than the width $\Delta$ of the ring resonances.  Under these conditions the phase $\chi$ will vary significantly over the BWF only as $N$ becomes a significant fraction of $N_{\text{coh}}$ \eqref{Ncohresult}.

\bibliography{SCISSOR_manuscript}

\begin{thebibliography}{37}
\expandafter\ifx\csname natexlab\endcsname\relax\def\natexlab#1{#1}\fi
\expandafter\ifx\csname bibnamefont\endcsname\relax
  \def\bibnamefont#1{#1}\fi
\expandafter\ifx\csname bibfnamefont\endcsname\relax
  \def\bibfnamefont#1{#1}\fi
\expandafter\ifx\csname citenamefont\endcsname\relax
  \def\citenamefont#1{#1}\fi
\expandafter\ifx\csname url\endcsname\relax
  \def\url#1{\texttt{#1}}\fi
\expandafter\ifx\csname urlprefix\endcsname\relax\def\urlprefix{URL }\fi
\providecommand{\bibinfo}[2]{#2}
\providecommand{\eprint}[2][]{\url{#2}}

\bibitem[{\citenamefont{Lanco et~al.}(2006)\citenamefont{Lanco, Ducci,
  Likforman, Marcadet, van Houwelingen, Zbinden, Leo, and Berger}}]{lanco06}
\bibinfo{author}{\bibfnamefont{L.}~\bibnamefont{Lanco}},
  \bibinfo{author}{\bibfnamefont{S.}~\bibnamefont{Ducci}},
  \bibinfo{author}{\bibfnamefont{J.-P.} \bibnamefont{Likforman}},
  \bibinfo{author}{\bibfnamefont{X.}~\bibnamefont{Marcadet}},
  \bibinfo{author}{\bibfnamefont{J.~A.~W.} \bibnamefont{van Houwelingen}},
  \bibinfo{author}{\bibfnamefont{H.}~\bibnamefont{Zbinden}},
  \bibinfo{author}{\bibfnamefont{G.}~\bibnamefont{Leo}}, \bibnamefont{and}
  \bibinfo{author}{\bibfnamefont{V.}~\bibnamefont{Berger}},
  \bibinfo{journal}{Phys. Rev. Lett.} \textbf{\bibinfo{volume}{97}},
  \bibinfo{pages}{173901} (\bibinfo{year}{2006}).

\bibitem[{\citenamefont{Takesue et~al.}(2007)\citenamefont{Takesue, Tokura,
  Fukuda, Tsuchizawa, Watanabe, Yamada, and Itabashi}}]{takesue07}
\bibinfo{author}{\bibfnamefont{H.}~\bibnamefont{Takesue}},
  \bibinfo{author}{\bibfnamefont{Y.}~\bibnamefont{Tokura}},
  \bibinfo{author}{\bibfnamefont{H.}~\bibnamefont{Fukuda}},
  \bibinfo{author}{\bibfnamefont{T.}~\bibnamefont{Tsuchizawa}},
  \bibinfo{author}{\bibfnamefont{T.}~\bibnamefont{Watanabe}},
  \bibinfo{author}{\bibfnamefont{K.}~\bibnamefont{Yamada}}, \bibnamefont{and}
  \bibinfo{author}{\bibfnamefont{S.-i.} \bibnamefont{Itabashi}},
  \bibinfo{journal}{Applied Physics Letters} \textbf{\bibinfo{volume}{91}},
  \bibinfo{eid}{201108} (\bibinfo{year}{2007}).

\bibitem[{\citenamefont{Clemmen et~al.}(2009)\citenamefont{Clemmen, Huy,
  Bogaerts, Baets, Emplit, and Massar}}]{clemmen09}
\bibinfo{author}{\bibfnamefont{S.}~\bibnamefont{Clemmen}},
  \bibinfo{author}{\bibfnamefont{K.~P.} \bibnamefont{Huy}},
  \bibinfo{author}{\bibfnamefont{W.}~\bibnamefont{Bogaerts}},
  \bibinfo{author}{\bibfnamefont{R.~G.} \bibnamefont{Baets}},
  \bibinfo{author}{\bibfnamefont{P.}~\bibnamefont{Emplit}}, \bibnamefont{and}
  \bibinfo{author}{\bibfnamefont{S.}~\bibnamefont{Massar}},
  \bibinfo{journal}{Opt. Express} \textbf{\bibinfo{volume}{17}},
  \bibinfo{pages}{16558} (\bibinfo{year}{2009}).

\bibitem[{\citenamefont{Xiong et~al.}(2011)\citenamefont{Xiong, Monat, Clark,
  Grillet, Marshall, Steel, Li, O'Faolain, Krauss, Rarity et~al.}}]{xiong11}
\bibinfo{author}{\bibfnamefont{C.}~\bibnamefont{Xiong}},
  \bibinfo{author}{\bibfnamefont{C.}~\bibnamefont{Monat}},
  \bibinfo{author}{\bibfnamefont{A.~S.} \bibnamefont{Clark}},
  \bibinfo{author}{\bibfnamefont{C.}~\bibnamefont{Grillet}},
  \bibinfo{author}{\bibfnamefont{G.~D.} \bibnamefont{Marshall}},
  \bibinfo{author}{\bibfnamefont{M.~J.} \bibnamefont{Steel}},
  \bibinfo{author}{\bibfnamefont{J.}~\bibnamefont{Li}},
  \bibinfo{author}{\bibfnamefont{L.}~\bibnamefont{O'Faolain}},
  \bibinfo{author}{\bibfnamefont{T.~F.} \bibnamefont{Krauss}},
  \bibinfo{author}{\bibfnamefont{J.~G.} \bibnamefont{Rarity}},
  \bibnamefont{et~al.}, \bibinfo{journal}{Opt. Lett.}
  \textbf{\bibinfo{volume}{36}}, \bibinfo{pages}{3413} (\bibinfo{year}{2011}).

\bibitem[{\citenamefont{Davan\c{c}o et~al.}(2012)\citenamefont{Davan\c{c}o,
  Ong, Shehata, Tosi, Agha, Assefa, Xia, Green, Mookherjea, and
  Srinivasan}}]{davanco12}
\bibinfo{author}{\bibfnamefont{M.}~\bibnamefont{Davan\c{c}o}},
  \bibinfo{author}{\bibfnamefont{J.~R.} \bibnamefont{Ong}},
  \bibinfo{author}{\bibfnamefont{A.~B.} \bibnamefont{Shehata}},
  \bibinfo{author}{\bibfnamefont{A.}~\bibnamefont{Tosi}},
  \bibinfo{author}{\bibfnamefont{I.}~\bibnamefont{Agha}},
  \bibinfo{author}{\bibfnamefont{S.}~\bibnamefont{Assefa}},
  \bibinfo{author}{\bibfnamefont{F.}~\bibnamefont{Xia}},
  \bibinfo{author}{\bibfnamefont{W.~M.~J.} \bibnamefont{Green}},
  \bibinfo{author}{\bibfnamefont{S.}~\bibnamefont{Mookherjea}},
  \bibnamefont{and}
  \bibinfo{author}{\bibfnamefont{K.}~\bibnamefont{Srinivasan}},
  \bibinfo{journal}{Applied Physics Letters} \textbf{\bibinfo{volume}{100}},
  \bibinfo{eid}{261104} (\bibinfo{year}{2012}).

\bibitem[{\citenamefont{Azzini et~al.}(2012{\natexlab{a}})\citenamefont{Azzini,
  Grassani, Strain, Sorel, Helt, Sipe, Liscidini, Galli, and
  Bajoni}}]{azzini12}
\bibinfo{author}{\bibfnamefont{S.}~\bibnamefont{Azzini}},
  \bibinfo{author}{\bibfnamefont{D.}~\bibnamefont{Grassani}},
  \bibinfo{author}{\bibfnamefont{M.~J.} \bibnamefont{Strain}},
  \bibinfo{author}{\bibfnamefont{M.}~\bibnamefont{Sorel}},
  \bibinfo{author}{\bibfnamefont{L.~G.} \bibnamefont{Helt}},
  \bibinfo{author}{\bibfnamefont{J.~E.} \bibnamefont{Sipe}},
  \bibinfo{author}{\bibfnamefont{M.}~\bibnamefont{Liscidini}},
  \bibinfo{author}{\bibfnamefont{M.}~\bibnamefont{Galli}}, \bibnamefont{and}
  \bibinfo{author}{\bibfnamefont{D.}~\bibnamefont{Bajoni}},
  \bibinfo{journal}{Opt. Express} \textbf{\bibinfo{volume}{20}},
  \bibinfo{pages}{23100} (\bibinfo{year}{2012}{\natexlab{a}}).

\bibitem[{\citenamefont{Ong and Mookherjea}(2013)}]{mookherjea13}
\bibinfo{author}{\bibfnamefont{J.~R.} \bibnamefont{Ong}} \bibnamefont{and}
  \bibinfo{author}{\bibfnamefont{S.}~\bibnamefont{Mookherjea}},
  \bibinfo{journal}{Opt. Express} \textbf{\bibinfo{volume}{21}},
  \bibinfo{pages}{5171} (\bibinfo{year}{2013}).

\bibitem[{\citenamefont{Orieux et~al.}(2013)\citenamefont{Orieux, Eckstein,
  Lema\^{i}tre, Filloux, Favero, Leo, Coudreau, Keller, Milman, and
  Ducci}}]{ducci13}
\bibinfo{author}{\bibfnamefont{A.}~\bibnamefont{Orieux}},
  \bibinfo{author}{\bibfnamefont{A.}~\bibnamefont{Eckstein}},
  \bibinfo{author}{\bibfnamefont{A.}~\bibnamefont{Lema\^{i}tre}},
  \bibinfo{author}{\bibfnamefont{P.}~\bibnamefont{Filloux}},
  \bibinfo{author}{\bibfnamefont{I.}~\bibnamefont{Favero}},
  \bibinfo{author}{\bibfnamefont{G.}~\bibnamefont{Leo}},
  \bibinfo{author}{\bibfnamefont{T.}~\bibnamefont{Coudreau}},
  \bibinfo{author}{\bibfnamefont{A.}~\bibnamefont{Keller}},
  \bibinfo{author}{\bibfnamefont{P.}~\bibnamefont{Milman}}, \bibnamefont{and}
  \bibinfo{author}{\bibfnamefont{S.}~\bibnamefont{Ducci}},
  \bibinfo{journal}{Phys. Rev. Lett.} \textbf{\bibinfo{volume}{110}},
  \bibinfo{pages}{160502} (\bibinfo{year}{2013}).

\bibitem[{\citenamefont{Streshinsky et~al.}(2013)\citenamefont{Streshinsky,
  Ding, Liu, Novack, Galland, Lim, Lo, Baehr-Jones, and Hochberg}}]{galland13}
\bibinfo{author}{\bibfnamefont{M.}~\bibnamefont{Streshinsky}},
  \bibinfo{author}{\bibfnamefont{R.}~\bibnamefont{Ding}},
  \bibinfo{author}{\bibfnamefont{Y.}~\bibnamefont{Liu}},
  \bibinfo{author}{\bibfnamefont{A.}~\bibnamefont{Novack}},
  \bibinfo{author}{\bibfnamefont{C.}~\bibnamefont{Galland}},
  \bibinfo{author}{\bibfnamefont{A.~E.-J.} \bibnamefont{Lim}},
  \bibinfo{author}{\bibfnamefont{P.~G.-Q.} \bibnamefont{Lo}},
  \bibinfo{author}{\bibfnamefont{T.}~\bibnamefont{Baehr-Jones}},
  \bibnamefont{and} \bibinfo{author}{\bibfnamefont{M.}~\bibnamefont{Hochberg}},
  \bibinfo{journal}{Opt. Photon. News} \textbf{\bibinfo{volume}{24}},
  \bibinfo{pages}{32} (\bibinfo{year}{2013}).

\bibitem[{\citenamefont{Kwiat et~al.}(1995)\citenamefont{Kwiat, Mattle,
  Weinfurter, Zeilinger, Sergienko, and Shih}}]{kwiat95}
\bibinfo{author}{\bibfnamefont{P.~G.} \bibnamefont{Kwiat}},
  \bibinfo{author}{\bibfnamefont{K.}~\bibnamefont{Mattle}},
  \bibinfo{author}{\bibfnamefont{H.}~\bibnamefont{Weinfurter}},
  \bibinfo{author}{\bibfnamefont{A.}~\bibnamefont{Zeilinger}},
  \bibinfo{author}{\bibfnamefont{A.~V.} \bibnamefont{Sergienko}},
  \bibnamefont{and} \bibinfo{author}{\bibfnamefont{Y.}~\bibnamefont{Shih}},
  \bibinfo{journal}{Phys. Rev. Lett.} \textbf{\bibinfo{volume}{75}},
  \bibinfo{pages}{4337} (\bibinfo{year}{1995}).

\bibitem[{\citenamefont{Lundeen and Steinberg}(2009)}]{steinberg09}
\bibinfo{author}{\bibfnamefont{J.~S.} \bibnamefont{Lundeen}} \bibnamefont{and}
  \bibinfo{author}{\bibfnamefont{A.~M.} \bibnamefont{Steinberg}},
  \bibinfo{journal}{Phys. Rev. Lett.} \textbf{\bibinfo{volume}{102}},
  \bibinfo{pages}{020404} (\bibinfo{year}{2009}).

\bibitem[{\citenamefont{Crespi et~al.}(2013)\citenamefont{Crespi, Osellame,
  Ramponi, Giovannetti, Fazio, Sansoni, De~Nicola, Sciarrino, and
  Mataloni}}]{mataloni13}
\bibinfo{author}{\bibfnamefont{A.}~\bibnamefont{Crespi}},
  \bibinfo{author}{\bibfnamefont{R.}~\bibnamefont{Osellame}},
  \bibinfo{author}{\bibfnamefont{R.}~\bibnamefont{Ramponi}},
  \bibinfo{author}{\bibfnamefont{V.}~\bibnamefont{Giovannetti}},
  \bibinfo{author}{\bibfnamefont{R.}~\bibnamefont{Fazio}},
  \bibinfo{author}{\bibfnamefont{L.}~\bibnamefont{Sansoni}},
  \bibinfo{author}{\bibfnamefont{F.}~\bibnamefont{De~Nicola}},
  \bibinfo{author}{\bibfnamefont{F.}~\bibnamefont{Sciarrino}},
  \bibnamefont{and} \bibinfo{author}{\bibfnamefont{P.}~\bibnamefont{Mataloni}},
  \bibinfo{journal}{Nature Photonics} \textbf{\bibinfo{volume}{7}},
  \bibinfo{pages}{322} (\bibinfo{year}{2013}), \bibinfo{note}{cited By 0}.

\bibitem[{\citenamefont{Spring et~al.}(2013)\citenamefont{Spring, Metcalf,
  Humphreys, Kolthammer, Jin, Barbieri, Datta, Thomas-Peter, Langford, Kundys
  et~al.}}]{walmsley13}
\bibinfo{author}{\bibfnamefont{J.~B.} \bibnamefont{Spring}},
  \bibinfo{author}{\bibfnamefont{B.~J.} \bibnamefont{Metcalf}},
  \bibinfo{author}{\bibfnamefont{P.~C.} \bibnamefont{Humphreys}},
  \bibinfo{author}{\bibfnamefont{W.~S.} \bibnamefont{Kolthammer}},
  \bibinfo{author}{\bibfnamefont{X.-M.} \bibnamefont{Jin}},
  \bibinfo{author}{\bibfnamefont{M.}~\bibnamefont{Barbieri}},
  \bibinfo{author}{\bibfnamefont{A.}~\bibnamefont{Datta}},
  \bibinfo{author}{\bibfnamefont{N.}~\bibnamefont{Thomas-Peter}},
  \bibinfo{author}{\bibfnamefont{N.~K.} \bibnamefont{Langford}},
  \bibinfo{author}{\bibfnamefont{D.}~\bibnamefont{Kundys}},
  \bibnamefont{et~al.}, \bibinfo{journal}{Science}
  \textbf{\bibinfo{volume}{339}}, \bibinfo{pages}{798} (\bibinfo{year}{2013}).

\bibitem[{\citenamefont{Broome et~al.}(2013)\citenamefont{Broome, Fedrizzi,
  Rahimi-Keshari, Dove, Aaronson, Ralph, and White}}]{white13}
\bibinfo{author}{\bibfnamefont{M.~A.} \bibnamefont{Broome}},
  \bibinfo{author}{\bibfnamefont{A.}~\bibnamefont{Fedrizzi}},
  \bibinfo{author}{\bibfnamefont{S.}~\bibnamefont{Rahimi-Keshari}},
  \bibinfo{author}{\bibfnamefont{J.}~\bibnamefont{Dove}},
  \bibinfo{author}{\bibfnamefont{S.}~\bibnamefont{Aaronson}},
  \bibinfo{author}{\bibfnamefont{T.~C.} \bibnamefont{Ralph}}, \bibnamefont{and}
  \bibinfo{author}{\bibfnamefont{A.~G.} \bibnamefont{White}},
  \bibinfo{journal}{Science} \textbf{\bibinfo{volume}{339}},
  \bibinfo{pages}{794} (\bibinfo{year}{2013}).

\bibitem[{\citenamefont{Tillmann et~al.}(2013)\citenamefont{Tillmann, Dakić,
  Heilmann, Nolte, Szameit, and Walther}}]{tilmann13}
\bibinfo{author}{\bibfnamefont{M.}~\bibnamefont{Tillmann}},
  \bibinfo{author}{\bibfnamefont{B.}~\bibnamefont{Dakić}},
  \bibinfo{author}{\bibfnamefont{R.}~\bibnamefont{Heilmann}},
  \bibinfo{author}{\bibfnamefont{S.}~\bibnamefont{Nolte}},
  \bibinfo{author}{\bibfnamefont{A.}~\bibnamefont{Szameit}}, \bibnamefont{and}
  \bibinfo{author}{\bibfnamefont{P.}~\bibnamefont{Walther}},
  \bibinfo{journal}{Nature Photonics} \textbf{\bibinfo{volume}{7}},
  \bibinfo{pages}{540} (\bibinfo{year}{2013}), \bibinfo{note}{cited By 0}.

\bibitem[{\citenamefont{Helt et~al.}(2012{\natexlab{a}})\citenamefont{Helt,
  Liscidini, and Sipe}}]{helt12}
\bibinfo{author}{\bibfnamefont{L.~G.} \bibnamefont{Helt}},
  \bibinfo{author}{\bibfnamefont{M.}~\bibnamefont{Liscidini}},
  \bibnamefont{and} \bibinfo{author}{\bibfnamefont{J.~E.} \bibnamefont{Sipe}},
  \bibinfo{journal}{J. Opt. Soc. Am. B} \textbf{\bibinfo{volume}{29}},
  \bibinfo{pages}{2199} (\bibinfo{year}{2012}{\natexlab{a}}).

\bibitem[{\citenamefont{Azzini et~al.}(2012{\natexlab{b}})\citenamefont{Azzini,
  Grassani, Galli, Andreani, Sorel, Strain, Helt, Sipe, Liscidini, and
  Bajoni}}]{azzini12_2}
\bibinfo{author}{\bibfnamefont{S.}~\bibnamefont{Azzini}},
  \bibinfo{author}{\bibfnamefont{D.}~\bibnamefont{Grassani}},
  \bibinfo{author}{\bibfnamefont{M.}~\bibnamefont{Galli}},
  \bibinfo{author}{\bibfnamefont{L.~C.} \bibnamefont{Andreani}},
  \bibinfo{author}{\bibfnamefont{M.}~\bibnamefont{Sorel}},
  \bibinfo{author}{\bibfnamefont{M.~J.} \bibnamefont{Strain}},
  \bibinfo{author}{\bibfnamefont{L.~G.} \bibnamefont{Helt}},
  \bibinfo{author}{\bibfnamefont{J.~E.} \bibnamefont{Sipe}},
  \bibinfo{author}{\bibfnamefont{M.}~\bibnamefont{Liscidini}},
  \bibnamefont{and} \bibinfo{author}{\bibfnamefont{D.}~\bibnamefont{Bajoni}},
  \bibinfo{journal}{Opt. Lett.} \textbf{\bibinfo{volume}{37}},
  \bibinfo{pages}{3807} (\bibinfo{year}{2012}{\natexlab{b}}).

\bibitem[{\citenamefont{Azzini et~al.}(2013)\citenamefont{Azzini, Grassani,
  Galli, Gerace, Patrini, Liscidini, Velha, and Bajoni}}]{azzini13}
\bibinfo{author}{\bibfnamefont{S.}~\bibnamefont{Azzini}},
  \bibinfo{author}{\bibfnamefont{D.}~\bibnamefont{Grassani}},
  \bibinfo{author}{\bibfnamefont{M.}~\bibnamefont{Galli}},
  \bibinfo{author}{\bibfnamefont{D.}~\bibnamefont{Gerace}},
  \bibinfo{author}{\bibfnamefont{M.}~\bibnamefont{Patrini}},
  \bibinfo{author}{\bibfnamefont{M.}~\bibnamefont{Liscidini}},
  \bibinfo{author}{\bibfnamefont{P.}~\bibnamefont{Velha}}, \bibnamefont{and}
  \bibinfo{author}{\bibfnamefont{D.}~\bibnamefont{Bajoni}},
  \bibinfo{journal}{Applied Physics Letters} \textbf{\bibinfo{volume}{103}},
  \bibinfo{eid}{031117} (\bibinfo{year}{2013}).

\bibitem[{\citenamefont{Absil et~al.}(2000)\citenamefont{Absil, Hryniewicz,
  Little, Cho, Wilson, Joneckis, and Ho}}]{absil00}
\bibinfo{author}{\bibfnamefont{P.~P.} \bibnamefont{Absil}},
  \bibinfo{author}{\bibfnamefont{J.~V.} \bibnamefont{Hryniewicz}},
  \bibinfo{author}{\bibfnamefont{B.~E.} \bibnamefont{Little}},
  \bibinfo{author}{\bibfnamefont{P.~S.} \bibnamefont{Cho}},
  \bibinfo{author}{\bibfnamefont{R.~A.} \bibnamefont{Wilson}},
  \bibinfo{author}{\bibfnamefont{L.~G.} \bibnamefont{Joneckis}},
  \bibnamefont{and} \bibinfo{author}{\bibfnamefont{P.-T.} \bibnamefont{Ho}},
  \bibinfo{journal}{Opt. Lett.} \textbf{\bibinfo{volume}{25}},
  \bibinfo{pages}{554} (\bibinfo{year}{2000}).

\bibitem[{\citenamefont{Dulkeith et~al.}(2006)\citenamefont{Dulkeith, Vlasov,
  Chen, Panoiu, and Richard M.~Osgood}}]{vlasov06}
\bibinfo{author}{\bibfnamefont{E.}~\bibnamefont{Dulkeith}},
  \bibinfo{author}{\bibfnamefont{Y.~A.} \bibnamefont{Vlasov}},
  \bibinfo{author}{\bibfnamefont{X.}~\bibnamefont{Chen}},
  \bibinfo{author}{\bibfnamefont{N.~C.} \bibnamefont{Panoiu}},
  \bibnamefont{and} \bibinfo{author}{\bibfnamefont{J.}~\bibnamefont{Richard
  M.~Osgood}}, \bibinfo{journal}{Opt. Express} \textbf{\bibinfo{volume}{14}},
  \bibinfo{pages}{5524} (\bibinfo{year}{2006}).

\bibitem[{\citenamefont{Yang et~al.}(2007)\citenamefont{Yang, Chak, Bristow,
  van Driel, Iyer, Aitchison, Smirl, and Sipe}}]{yang07}
\bibinfo{author}{\bibfnamefont{Z.}~\bibnamefont{Yang}},
  \bibinfo{author}{\bibfnamefont{P.}~\bibnamefont{Chak}},
  \bibinfo{author}{\bibfnamefont{A.~D.} \bibnamefont{Bristow}},
  \bibinfo{author}{\bibfnamefont{H.~M.} \bibnamefont{van Driel}},
  \bibinfo{author}{\bibfnamefont{R.}~\bibnamefont{Iyer}},
  \bibinfo{author}{\bibfnamefont{J.~S.} \bibnamefont{Aitchison}},
  \bibinfo{author}{\bibfnamefont{A.~L.} \bibnamefont{Smirl}}, \bibnamefont{and}
  \bibinfo{author}{\bibfnamefont{J.~E.} \bibnamefont{Sipe}},
  \bibinfo{journal}{Opt. Lett.} \textbf{\bibinfo{volume}{32}},
  \bibinfo{pages}{826} (\bibinfo{year}{2007}).

\bibitem[{\citenamefont{Turner et~al.}(2008)\citenamefont{Turner, Foster,
  Gaeta, and Lipson}}]{turner08}
\bibinfo{author}{\bibfnamefont{A.~C.} \bibnamefont{Turner}},
  \bibinfo{author}{\bibfnamefont{M.~A.} \bibnamefont{Foster}},
  \bibinfo{author}{\bibfnamefont{A.~L.} \bibnamefont{Gaeta}}, \bibnamefont{and}
  \bibinfo{author}{\bibfnamefont{M.}~\bibnamefont{Lipson}},
  \bibinfo{journal}{Opt. Express} \textbf{\bibinfo{volume}{16}},
  \bibinfo{pages}{4881} (\bibinfo{year}{2008}).

\bibitem[{\citenamefont{Corcoran et~al.}(2009)\citenamefont{Corcoran, Monat,
  Grillet, Moss, Eggleton, White, O'Faolain, and Krauss}}]{corcoran09}
\bibinfo{author}{\bibfnamefont{B.}~\bibnamefont{Corcoran}},
  \bibinfo{author}{\bibfnamefont{C.}~\bibnamefont{Monat}},
  \bibinfo{author}{\bibfnamefont{C.}~\bibnamefont{Grillet}},
  \bibinfo{author}{\bibfnamefont{D.}~\bibnamefont{Moss}},
  \bibinfo{author}{\bibfnamefont{B.}~\bibnamefont{Eggleton}},
  \bibinfo{author}{\bibfnamefont{T.}~\bibnamefont{White}},
  \bibinfo{author}{\bibfnamefont{L.}~\bibnamefont{O'Faolain}},
  \bibnamefont{and} \bibinfo{author}{\bibfnamefont{T.}~\bibnamefont{Krauss}},
  \bibinfo{journal}{Nature Photonics} \textbf{\bibinfo{volume}{3}},
  \bibinfo{pages}{206} (\bibinfo{year}{2009}), \bibinfo{note}{cited By 0}.

\bibitem[{\citenamefont{Galli et~al.}(2010)\citenamefont{Galli, Gerace, Welna,
  Krauss, O'Faolain, Guizzetti, and Andreani}}]{galli10}
\bibinfo{author}{\bibfnamefont{M.}~\bibnamefont{Galli}},
  \bibinfo{author}{\bibfnamefont{D.}~\bibnamefont{Gerace}},
  \bibinfo{author}{\bibfnamefont{K.}~\bibnamefont{Welna}},
  \bibinfo{author}{\bibfnamefont{T.~F.} \bibnamefont{Krauss}},
  \bibinfo{author}{\bibfnamefont{L.}~\bibnamefont{O'Faolain}},
  \bibinfo{author}{\bibfnamefont{G.}~\bibnamefont{Guizzetti}},
  \bibnamefont{and} \bibinfo{author}{\bibfnamefont{L.~C.}
  \bibnamefont{Andreani}}, \bibinfo{journal}{Opt. Express}
  \textbf{\bibinfo{volume}{18}}, \bibinfo{pages}{26613} (\bibinfo{year}{2010}).

\bibitem[{\citenamefont{Helt et~al.}(2012{\natexlab{b}})\citenamefont{Helt,
  Sipe, and Liscidini}}]{helt12_2}
\bibinfo{author}{\bibfnamefont{L.~G.} \bibnamefont{Helt}},
  \bibinfo{author}{\bibfnamefont{J.~E.} \bibnamefont{Sipe}}, \bibnamefont{and}
  \bibinfo{author}{\bibfnamefont{M.}~\bibnamefont{Liscidini}},
  \bibinfo{journal}{Opt. Lett.} \textbf{\bibinfo{volume}{37}},
  \bibinfo{pages}{4431} (\bibinfo{year}{2012}{\natexlab{b}}).

\bibitem[{\citenamefont{E.~Heebner et~al.}(2002)\citenamefont{E.~Heebner, Boyd,
  and Park}}]{heebner02}
\bibinfo{author}{\bibfnamefont{J.}~\bibnamefont{E.~Heebner}},
  \bibinfo{author}{\bibfnamefont{R.~W.} \bibnamefont{Boyd}}, \bibnamefont{and}
  \bibinfo{author}{\bibfnamefont{Q.-H.} \bibnamefont{Park}},
  \bibinfo{journal}{Phys. Rev. E} \textbf{\bibinfo{volume}{65}},
  \bibinfo{pages}{036619} (\bibinfo{year}{2002}).

\bibitem[{\citenamefont{Heebner et~al.}(2004)\citenamefont{Heebner, Chak,
  Pereira, Sipe, and Boyd}}]{heebner04}
\bibinfo{author}{\bibfnamefont{J.~E.} \bibnamefont{Heebner}},
  \bibinfo{author}{\bibfnamefont{P.}~\bibnamefont{Chak}},
  \bibinfo{author}{\bibfnamefont{S.}~\bibnamefont{Pereira}},
  \bibinfo{author}{\bibfnamefont{J.~E.} \bibnamefont{Sipe}}, \bibnamefont{and}
  \bibinfo{author}{\bibfnamefont{R.~W.} \bibnamefont{Boyd}},
  \bibinfo{journal}{J. Opt. Soc. Am. B} \textbf{\bibinfo{volume}{21}},
  \bibinfo{pages}{1818} (\bibinfo{year}{2004}).

\bibitem[{\citenamefont{Dicke}(1954)}]{dicke54}
\bibinfo{author}{\bibfnamefont{R.~H.} \bibnamefont{Dicke}},
  \bibinfo{journal}{Phys. Rev.} \textbf{\bibinfo{volume}{93}},
  \bibinfo{pages}{99} (\bibinfo{year}{1954}).

\bibitem[{\citenamefont{Liscidini et~al.}(2012)\citenamefont{Liscidini, Helt,
  and Sipe}}]{liscidini12}
\bibinfo{author}{\bibfnamefont{M.}~\bibnamefont{Liscidini}},
  \bibinfo{author}{\bibfnamefont{L.~G.} \bibnamefont{Helt}}, \bibnamefont{and}
  \bibinfo{author}{\bibfnamefont{J.~E.} \bibnamefont{Sipe}},
  \bibinfo{journal}{Phys. Rev. A} \textbf{\bibinfo{volume}{85}},
  \bibinfo{pages}{013833} (\bibinfo{year}{2012}).

\bibitem[{\citenamefont{Breit and Bethe}(1954)}]{breit54}
\bibinfo{author}{\bibfnamefont{G.}~\bibnamefont{Breit}} \bibnamefont{and}
  \bibinfo{author}{\bibfnamefont{H.~A.} \bibnamefont{Bethe}},
  \bibinfo{journal}{Phys. Rev.} \textbf{\bibinfo{volume}{93}},
  \bibinfo{pages}{888} (\bibinfo{year}{1954}).

\bibitem[{\citenamefont{Agrawal}(2007)}]{agrawal07}
\bibinfo{author}{\bibfnamefont{G.~P.} \bibnamefont{Agrawal}},
  \emph{\bibinfo{title}{Nonlinear Fiber Optics}} (\bibinfo{publisher}{Academic
  Press}, \bibinfo{address}{Burlington, MA}, \bibinfo{year}{2007}),
  \bibinfo{edition}{4th} ed.

\bibitem[{\citenamefont{Sipe et~al.}(2004)\citenamefont{Sipe, Bhat, Chak, and
  Pereira}}]{sipe04}
\bibinfo{author}{\bibfnamefont{J.~E.} \bibnamefont{Sipe}},
  \bibinfo{author}{\bibfnamefont{N.~A.~R.} \bibnamefont{Bhat}},
  \bibinfo{author}{\bibfnamefont{P.}~\bibnamefont{Chak}}, \bibnamefont{and}
  \bibinfo{author}{\bibfnamefont{S.}~\bibnamefont{Pereira}},
  \bibinfo{journal}{Phys. Rev. E} \textbf{\bibinfo{volume}{69}},
  \bibinfo{pages}{016604} (\bibinfo{year}{2004}).

\bibitem[{\citenamefont{Quesada and Sipe}(2015)}]{Quesada2015}
\bibinfo{author}{\bibfnamefont{N.}~\bibnamefont{Quesada}} \bibnamefont{and}
  \bibinfo{author}{\bibfnamefont{J.~E.} \bibnamefont{Sipe}},
  \bibinfo{journal}{Phys. Rev. Lett.} \textbf{\bibinfo{volume}{114}},
  \bibinfo{pages}{093903} (\bibinfo{year}{2015}).

\bibitem[{\citenamefont{Christ et~al.}(2011)\citenamefont{Christ, Laiho,
  Eckstein, Cassemiro, and Silberhorn}}]{christ11}
\bibinfo{author}{\bibfnamefont{A.}~\bibnamefont{Christ}},
  \bibinfo{author}{\bibfnamefont{K.}~\bibnamefont{Laiho}},
  \bibinfo{author}{\bibfnamefont{A.}~\bibnamefont{Eckstein}},
  \bibinfo{author}{\bibfnamefont{K.~N.} \bibnamefont{Cassemiro}},
  \bibnamefont{and}
  \bibinfo{author}{\bibfnamefont{C.}~\bibnamefont{Silberhorn}},
  \bibinfo{journal}{New Journal of Physics} \textbf{\bibinfo{volume}{13}},
  \bibinfo{pages}{033027} (\bibinfo{year}{2011}).

\bibitem[{\citenamefont{Heebner et~al.}(2008)\citenamefont{Heebner, Grover, and
  Ibrahim}}]{heebner08}
\bibinfo{author}{\bibfnamefont{J.~E.} \bibnamefont{Heebner}},
  \bibinfo{author}{\bibfnamefont{R.}~\bibnamefont{Grover}}, \bibnamefont{and}
  \bibinfo{author}{\bibfnamefont{T.}~\bibnamefont{Ibrahim}},
  \emph{\bibinfo{title}{Optical Microresonators: Theory, Fabrication, and
  Applications}} (\bibinfo{publisher}{Springer}, \bibinfo{address}{London},
  \bibinfo{year}{2008}).

\bibitem[{\citenamefont{Helt et~al.}(2011)\citenamefont{Helt, Liscidini, Farsi,
  Clemmen, Venkataraman, Levy, Lipson, Gaeta, and Sipe}}]{helt11}
\bibinfo{author}{\bibfnamefont{L.~G.} \bibnamefont{Helt}},
  \bibinfo{author}{\bibfnamefont{M.}~\bibnamefont{Liscidini}},
  \bibinfo{author}{\bibfnamefont{A.}~\bibnamefont{Farsi}},
  \bibinfo{author}{\bibfnamefont{S.}~\bibnamefont{Clemmen}},
  \bibinfo{author}{\bibfnamefont{V.}~\bibnamefont{Venkataraman}},
  \bibinfo{author}{\bibfnamefont{J.~S.} \bibnamefont{Levy}},
  \bibinfo{author}{\bibfnamefont{M.}~\bibnamefont{Lipson}},
  \bibinfo{author}{\bibfnamefont{A.~L.} \bibnamefont{Gaeta}}, \bibnamefont{and}
  \bibinfo{author}{\bibfnamefont{J.~E.} \bibnamefont{Sipe}}, in
  \emph{\bibinfo{booktitle}{CLEO:2011 - Laser Applications to Photonic
  Applications}} (\bibinfo{publisher}{Optical Society of America},
  \bibinfo{year}{2011}), p. \bibinfo{pages}{QWA4}.

\bibitem[{\citenamefont{Yang et~al.}(2008)\citenamefont{Yang, Liscidini, and
  Sipe}}]{yang08}
\bibinfo{author}{\bibfnamefont{Z.}~\bibnamefont{Yang}},
  \bibinfo{author}{\bibfnamefont{M.}~\bibnamefont{Liscidini}},
  \bibnamefont{and} \bibinfo{author}{\bibfnamefont{J.~E.} \bibnamefont{Sipe}},
  \bibinfo{journal}{Phys. Rev. A} \textbf{\bibinfo{volume}{77}},
  \bibinfo{pages}{033808} (\bibinfo{year}{2008}).

\end{thebibliography}

\end{document}